\documentclass[11pt,preprint]{aastex}
\usepackage{graphics}
\usepackage{subfigure}
\usepackage{hyperref}

\usepackage{natbib}
\usepackage{float}
\newcommand{\CH}[1]{\colhead{#1}}
\newcommand{\D}{$^{\dagger}$}
\newcommand{\st}{$^{\star}$}
\newcommand{\ds}{$^{\dagger \star}$}

\shortauthors{Berg et al.}
\title{New Radial Abundance Gradients for NGC 628 and NGC 2403}

\author{Danielle A. Berg\altaffilmark{1}, Evan D. Skillman\altaffilmark{1}, Donald R. Garnett, Kevin V. Croxall\altaffilmark{2},  
Andrew R. Marble\altaffilmark{3}, J.D. Smith\altaffilmark{4}, Karl Gordon\altaffilmark{5}, Robert C. Kennicutt, Jr.\altaffilmark{6,7}}

\altaffiltext{1}{Department of Astronomy, University of Minnesota, 116 Church St. SE, Minneapolis, MN 55455; berg@astro.umn.edu; skillman@astro.umn.edu}
\altaffiltext{2}{Ohio State University, 140 West 18th Avenue, Columbus, OH 43210; croxall.5@osu.edu}
\altaffiltext{3}{National Solar Observatory, 950 N Cherry Ave, Tucson, AZ 85719; amarble@nso.edu}
\altaffiltext{4}{Ritter Astrophysical Observatory, University of Toledo, Toledo, OH 43606, USA; jd.smith@utoledo.edu}
\altaffiltext{5}{Space Telescope Science Institute, 3700 San Martin Drive, Baltimore, MD 21218, USA; kgordon@stsci.edu}
\altaffiltext{6}{Institute of Astronomy, University of Cambridge, Madingley Road, Cambridge CB3 0HA, UK; robk@ast.cam.ac.uk}
\altaffiltext{7}{Steward Observatory, University of Arizona, Tucson, AZ 85721}

\begin{document}
\begin{abstract}

\end{abstract}

Motived by recent ISM studies, we present high quality MMT and 
Gemini spectroscopic observations of \ion{H}{2} regions 
in the nearby spiral galaxies NGC 628 and NGC 2403 in 
order to measure their chemical abundance gradients.
Using long-slit and multi-object mask optical spectroscopy, 
we obtained measurements of the temperature sensitive 
auroral lines [\ion{O}{3}] $\lambda4363$ and/or [\ion{N}{2}] 
$\lambda5755$ at a strength of 4$\sigma$ or greater in 11 \ion{H}{2} 
regions in NGC 628 and 7 regions in NGC 2403.
These observations allow us, for the first time, to derive 
an oxygen abundance gradient in NGC 628 based solely
on ``direct" oxygen abundances of \ion{H}{2} regions:
12 + log(O/H) = (8.43$\pm$0.03) + (-0.017$\pm$0.002)$\times{R_g}$ (dex/kpc),
with a dispersion in log(O/H) of $\sigma$ = 0.10 dex, 
from 14 regions with a radial coverage of $\sim$2-19 kpc. 
This is a significantly shallower slope than found by 
previous ``strong-line" abundance studies.
In NGC 2403, we derive an oxygen abundance gradient of 
12 + log(O/H) = (8.48$\pm$0.04) + (-0.032$\pm$0.007)$\times{R_g}$ (dex/kpc),
with a dispersion in log(O/H) of $\sigma$ = 0.07 dex,
 from 7 \ion{H}{2} with a radial coverage  of $\sim$1-10 kpc.

Additionally, we measure the N, S, Ne, and Ar abundances.
We find the N/O ratio decreases with increasing radius for the inner disk, 
but reaches a plateau past $R_{25}$ in NGC 628.
NGC 2403 also has a negative N/O gradient with radius, but we do not sample the 
outer disk of the galaxy past $R_{25}$ and so do not see evidence for a plateau.
This bi-modal pattern measured for NGC 628 indicates dominant contributions from secondary 
nitrogen inside of the $R_{25}$ transition and dominantly primary nitrogen farther out. 
As expected for $\alpha$-process elements, S/O, Ne/O, and Ar/O are 
consistent with constant values over a range in oxygen abundance.

\keywords{galaxies: abundances - galaxies: spiral - galaxies: evolution}

%----------------------------------------------------------------------------------------------------------------------
%----------------------------------------------------------------------------------------------------------------------

\section{INTRODUCTION}\label{sec:intro}

\ion{H}{2} regions can be used to study absolute and relative 
abundances in the interstellar medium (ISM) of galaxies.
\citet{aller42} and \citet{searle71} were the first to infer radial 
gradients in excitation across the disks of spiral galaxies.
Since then, numerous studies have shown that, typically, spiral 
galaxies have radial abundance gradients in the sense of decreasing 
absolute abundances with increasing galactocentric radius.
This trend was first measured in our own galaxy by \citet{shaver83} 
and was confirmed by observations of other nearby spiral galaxies 
\citep[e.g.,][]{pagel81,shields90}. 
Thus, spiral galaxies in the nearby universe with low inclinations offer 
the opportunity to measure chemical abundances and subsequently 
compare these abundances with variations in physical conditions.

The dust properties of the ISM in spiral galaxies (such as the dust-to-gas ratio 
and the abundance of polycyclic aromatic hydrocarbons (PAHs)) 
are known to show systematic radial variations.
Previous studies have investigated the correlation between PAH emission
and metallicity, and found that PAH emission drops below a critical metallicity 
of 12+log(O/H)$\approx$8.0 in nearby dwarf galaxies
\citep[e.g.,][]{engelbracht05,engelbracht08,marble10}.
These works suggest that metallicity may play a role in PAH processing, and thus PAH 
abundance relative to total dust content, but other factors such as star formation 
and/or the local radiation field density affect the excitation of those molecules.
Other studies have supported the hypothesis that grain formation and modification is 
affected by local metallicity \citep[e.g.,][]{smith07,sandstrom12}.

Whether or not this result is universal is key to understanding PAH 
behavior in the ISM and requires a greater pool of reliable data.
\citet{smith05} carried out a successful \textit{Spitzer} observing campaign with deep, 
spatially resolved, low-resolution spectral maps from 5 to 38 microns in three nearby 
galaxies (M~101, NGC~628, and NGC~2403) to address these questions. 
Since the physical conditions in the ISM are likely important to the star formation process
and play a role in the composition, conditions, and size distribution of dust grains, 
understanding the connection between chemical abundances and the physical conditions in 
the ISM is fundamental to understanding star formation and galaxy evolution.  
Thus we are motivated to measure reliable direct abundance gradients
 for NGC~628 and NGC~2403; these results are being incorporated
 into the PAH studies of \citet{sandstrom13} and \citet{smith13}.

Detailed direct abundance studies exist for only a handful of spiral galaxies.
For example, direct \ion{H}{2} region abundances have been successfully measured 
for 20 high signal-to-noise \ion{H}{2} region spectra in M~101 by \citet{kennicutt03a}.
However, only two \ion{H}{2} region auroral line detections exist in the literature for NGC~628 
\citep{castellanos02}, and so no direct oxygen abundance gradient has ever been reported.
The \ion{H}{2} regions of NGC~2403 have been studied by \citet{garnett97},
but there were concerns about the linearity of the detector used.

In order to place gradients on the same scale for comparison amongst 
galaxies, reliable and consistent derivations of abundances are needed 
\citep[see, e.g.,][]{moustakas10}.
However, the vast majority of nebular abundance measurements for \ion{H}{2} 
regions in spiral galaxies are based upon ``strong-line'' observations which lack 
a ``direct" measurement of the electron temperature in the ionized gas.  
The conversion of these strong-line observations
into chemical abundances can be very uncertain and biased 
\citep{kennicutt03a, bresolin07, yin07, perez-montero09, berg11},
limiting their ability to provide meaningful comparisons between different galaxies.
Therefore, a ``direct" measurement of the electron temperature - 
typically derived from the ratio of auroral to collisionally excited lines - 
is needed for each \ion{H}{2} region.
Throughout the remainder of the paper, abundances determined 
in this fashion will be referred to as direct abundances.
Since the auroral lines become exponentially weaker with decreasing 
temperature (increasing abundance), observations become increasingly 
challenging at higher metallicities (small radii in spiral galaxies).
Nonetheless, with large telescope apertures and efficient spectrographs, 
it is possible to determine accurate chemical abundance gradients for spiral 
galaxies \citep[e.g.,][]{kennicutt03a, garnett04, bresolin04, bresolin07, 
bresolin09a, bresolin09b, bresolin11, zurita12}.
Note that direct chemical abundances are not without problems (they assume 
a uniform temperature distribution, ignoring temperature fluctuations) 
and great progress is being made to overcome their limitations 
\citep[e.g.,][]{esteban09,pena-guerrero12,nicholls12}.
Even so, direct abundances do provide a stable and well understood 
scale by which to make comparisons between galaxies.

A major unresolved issue in \ion{H}{2} region abundance studies is the 
importance of primary versus secondary production of nitrogen. 
Oxygen production is generally understood to be dominated by primary 
nucleosynthesis from massive stars and delivered early after a star formation event.
Like oxygen, nitrogen can be produced by massive stars (and delivered early with 
the primary oxygen), as well as by intermediate mass stars (and delivered relatively later). 
Nitrogen can have both primary and secondary origins.
\citet{garnett90} found a relatively constant relationship in the N/O ratio versus O/H 
for low metallicity star-forming galaxies.
At higher metallicities (i.e., 12+log(O/H) $\ge$ 8.0), secondary 
nitrogen production becomes increasingly significant, 
causing the average N/O to increase with O/H \citep{pagel85}.
We can, therefore, use the radial relationship of N/O in spiral galaxies 
to determine which nucleosynthetic mechanisms are dominant.

A further goal of \ion{H}{2} region spectroscopic studies it to measure
$\alpha$-element abundances as an observational constraint of
the IMF and stellar nucleosynthesis models.
Sulfur, neon, and argon are all $\alpha$-process elements which 
are produced through hydrostatic burning and explosive 
nucleosynthesis: neon is a product of carbon burning, 
while sulfur and argon are produced during oxygen burning.
In a comprehensive study of M33, \citet{kwitter81} found that Ne, 
N, S, and Ar gradients followed that which they derived for oxygen. 
The stellar nucleosynthesis calculations of \citet{woosley95} modeled these trends, 
indicating the $\alpha$-elements and oxygen are produced mainly in 
massive stars in a small mass range, and thus are expected to trace each other closely.
In contrast to this viewpoint, \citet{willner02} derived a neon gradient that 
is significantly shallower than the oxygen gradient observed in M33. 
While sulfur is also traditionally assumed to have a constant S/O ratio 
\citep{garnett89}, some specific cases, such as the work of \citet{vilchez88} 
on M33, find a slower decline of sulfur than oxygen with radius.
Thus additional $\alpha$-element observations are needed to properly
constrain stellar nucleosynthesis models.

Here we present new MMT and Gemini observations of \ion{H}{2} regions in NGC~628 and NGC~2403, 
which allow improved measurements of their chemical abundance gradients.
These allow us to estimate the abundance gradient of NGC~628 for the first time 
solely from direct abundances, and to improve the abundance gradient of NGC~2403 
by increasing the number of \ion{H}{2} regions with direct abundance measurements.
In Section~\ref{data} we describe  the MMT and Gemini observations 
and how we processed the spectra.
Emission line measurements and abundance determinations are detailed in 
\S~\ref{sec:abund}, which allows the discussion of the oxygen abundance 
gradient in \S~\ref{sec:O}, the nitrogen abundance gradient in \S~\ref{sec:N},
and of the sulfur, neon, and argon abundances in \S~\ref{sec:alpha}.
Finally, we summarize our findings in \S~\ref{sec:conclusion}. 

%----------------------------------------------------------------------------------------------------------------------
%----------------------------------------------------------------------------------------------------------------------

\section{NEW SPECTROSCOPIC OBSERVATIONS}\label{data}

%----------------------------------------------------------------------------------------------------------------------

\subsection{NGC~628 Spectra}\label{sec:mmt628}
NGC~628 (M~74) is a late-type giant spiral ScI galaxy with a 
systematic velocity of 656 km s$^{-1}$. 
We adopt a distance of 7.2 Mpc \citep{vandyk06} and an inclination of 
$i\approx5^{\circ}$ \citep{shostak84}, with a resulting scale of 35 pc arcsec$^{-1}$.
The optical parameters of NGC~628 are listed in Table~\ref{tbl0}.
NGC~628 is an excellent target due to its small inclination, 
extended structure, and undisturbed optical profile.
The gas-phase oxygen abundance of NGC~628 has been previously studied 
using long-slit spectroscopy \citep[e.g.,][]{talent83,mccall85,zaritsky94,ferguson98,
vanzee98b,bresolin99,castellanos02,moustakas10,gusev12,cedres12}, 
and integral field spectroscopy \citep[e.g.,][]{rosales-ortega11}.
Using strong-line abundances, these studies mostly found a 
constant gradient out to $R_g\sim1.7\cdot R_{25}$.
In contrast, \citet{rosales-ortega11} found a trimodal oxygen abundance 
gradient, where the innermost distribution is nearly flat, followed by a 
steep negative gradient out to $R_{25}$, and another nearly constant 
gradient beyond the optical edge of the galaxy.
Additionally, they found that the slope of the gradient depends strongly 
on the abundance calibrator used, concluding that this may be due to 
the potential for [\ion{N}{2}]-based empirical indices to overestimate oxygen 
abundance at high N/O ratios and vice versa \citep[see e.g.,][]{perez-montero09}.

New MMT and Gemini observations were acquired in order to achieve high signal-to-noise 
(S$/$N) spectra with the goal of detecting the faint [\ion{O}{3}] $\lambda$4363 
or [\ion{N}{2}] $\lambda$5755 auroral lines at a strength of 4$\sigma$ or higher. 
The Gemini observations were obtained from two multi-slit fields covering the inner parts of
NGC~628 and these were supplemented by MMT observations of individual \ion{H}{2} regions
in the outer parts of NGC~628.

Intermediate-resolution spectra of \ion{H}{2} regions in NGC 628 were obtained 
with the Gemini Multi-Object Spectrographs \citep[GMOS;][]{hook04} 
on the UT dates of 2006 September 21 and November 17.
The multi-object mode of GMOS, which uses custom-designed, laser-milled masks,
offers the possibility of obtaining spectra of many \ion{H}{2} regions simultaneously. 
Pre-imaging in an H$\alpha$ filter was used to identify \ion{H}{2} regions and 
determine accurate astrometry for the masks.
\ion{H}{2} regions were selected based on high H$\alpha$ surface brightness and
a large radial coverage of the disk.

Two masks were observed in queue mode, one of the North-West corner 
of NGC 628, and one placed on the South-East corner.
The NW and SE masks contained slits 1.5\arcsec\ wide covering 7 
and 8 different \ion{H}{2} regions respectively.
Slit lengths varied between 15-50\arcsec\ depending on the the size of the 
targeted \ion{H}{2} region and the proximity of other slits on the mask.  
Both blue and red spectra were obtained using a 600 line grating, 
giving a resolution of 0.45 \AA\ per pixel and a full width half maximum 
resolution of $\approx6$ \AA.
The red side additionally used the GG455{\textunderscore}G0305 order blocking filter.
Typical exposure times were 3$\times$1675 
seconds with the mask fixed to a position angle of zero, and the observations 
were obtained near transit.
Observations were centered at 4600 \AA, 4625 \AA, and 4650 \AA\ in the blue, 
and 6100 \AA, 6125 \AA, and 6150 \AA\ in the red, to ensure full 
spectral coverage across the detector gaps.

The NGC 628 MMT observations were obtained with the Blue Channel 
spectrograph \citep{schmidt89} on the UT dates of 2008 October 30-November 
1, 2009 June 15-22, and 2010 January 11-12. 
Sky conditions varied, but data were only acquired during 
minimal cloud coverage and approximately arcsecond seeing.
A 500 line grating, $1\arcsec$ slit, and UV-36 blocking filter were used, 
yielding an approximate dispersion of 1.2 \AA\ per pixel, a full width 
at half maximum resolution of $\lesssim3$ \AA, and a wavelength 
coverage of 3690--6790 \AA. 
The MMT and Blue Channel spectrograph combination provided the 
balance between sensitivity, resolution, and wavelength coverage 
conditions necessary to measure all emission lines relevant to 
oxygen abundance determinations.
Bias frames, flat-field lamp images, and sky flats were taken each night. 
Multiple standard stars from \citet{oke90} with spectral energy distributions 
peaking in the blue and containing minimal absorption were observed 
throughout the night using a 5$\arcsec$ slit over a range of airmasses. 

In each \ion{H}{2} region, the slit center was aligned with H$\alpha$ 
emission such that the surface brightness over the area of the slit 
was maximized.
Typically, three 1200 or 1800 second exposures were made for 
the MMT observations, with the slit at a fixed position angle which 
approximated the parallactic angle at the midpoint of the observation.
This, in addition to observing the galaxies at airmasses less than 1.5, 
served to minimize the wavelength-dependent light loss due to 
differential refraction \citep{filippenko82}. 

\begin{figure}[H]
\epsscale{1.0}
\figurenum{1}
\plotone{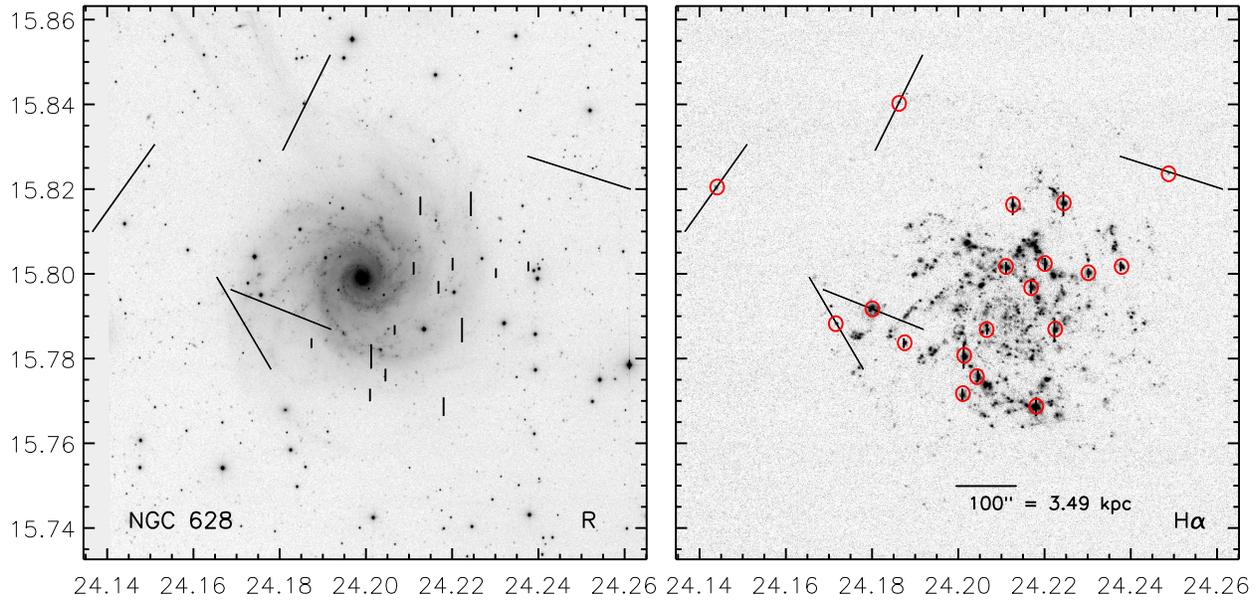}
\caption{R-band and continuum-subtracted H$\alpha$ images of NGC 628 (van Zee et al., in prep.).
The lines represent the various slit positions observed at Gemini and the MMT. 
The central slit positions targeted \ion{H}{2} regions and are indicated with red circles. 
See Tables~\ref{tbl1} and \ref{tbl2} for more details.}
\label{fig1}
\end{figure}

Finally, combined helium, argon, and neon arc lamps were observed at 
each pointing at the MMT, while copper-argon arc lamps were obtained at 
GEMINI for accurate wavelength calibration.
Tables~\ref{tbl1} and \ref{tbl2} list the log information for both 
the GMOS and MMT observations.
Figure~\ref{fig1} shows the R-band continuum and H$\alpha$ 
continuum-subtracted images for NGC 628 (van Zee et al., in prep.), with slit positions from 
Tables~\ref{tbl1} and \ref{tbl2} shown with black lines centered on red circles. 
The slit centers are marked in the H$\alpha$ image by red circles.

\subsection{NGC~2403 Spectra}\label{sec:mmt2403}
As a bright, nearby galaxy with a favorable inclination, NGC~2403 
allows observations of the chemical composition throughout its disk.
NGC 2403 is an intermediate luminosity, isolated spiral SABcd galaxy in the M81 Group.
We adopt a distance of 3.16 Mpc \citep{jacobs09} and an inclination 
of $\approx60^{\circ}$ with a resulting scale of 15.3 pc arcsec$^{-1}$. 
The optical parameters of NGC~2403 are listed in Table~\ref{tbl0}.
Several strong-line spectroscopic studies relevant to our work exist 
for NGC 2403 \citep[e.g.,][]{mccall85,fierro86,vanzee98b}.
In contrast, \citet{garnett97} measured direct abundances using the [\ion{O}{3}] 
$\lambda4363$, [\ion{S}{3}] $\lambda6312$, and [\ion{O}{2}] $\lambda$7320-7330 
emission lines to determine the electron temperatures for 9 \ion{H}{2} regions.
Although the [\ion{O}{3}] $\lambda$4363 line was detected at high confidence in these
regions, there were some lingering concerns about the effects of the non-linearity
of the IPCS detector that appear at moderately high count rates \citep{jenkins87}.
Thus, checking on the reliability of the previous observations and increasing the 
number of measurements and radial coverage is desirable.

New MMT observations were acquired in order to achieve high signal-to-noise 
(S$/$N) spectra with the goal of measuring direct abundances. 
Observations for NGC 2403 were acquired using the Blue Channel 
Spectrograph at the MMT on the UT dates of 2006 February 1-4.
A setup similar to the NGC 628 observations was used, but the exposure 
time was varied corresponding to the surface brightness of the region.
A 500 mm$^{-1}$ grating was used providing a 3650-6790 \AA\ coverage.
The \ion{H}{2} regions of NGC~2403 have been identified and cataloged 
by \citet{veron65}, \citet{hodge83}, and \citet{sivan90}.
Targets were chosen to overlap with the sample from \citet[][here after 
referred to as G97]{garnett97} and also to improve the radial coverage.
For five of the targets, a 832 mm$^{-1}$ grating was also used,
providing additional wavelength coverage from 5495 to 7405 \AA.

\begin{figure}[H]
\epsscale{1.0}
\figurenum{2}
\plotone{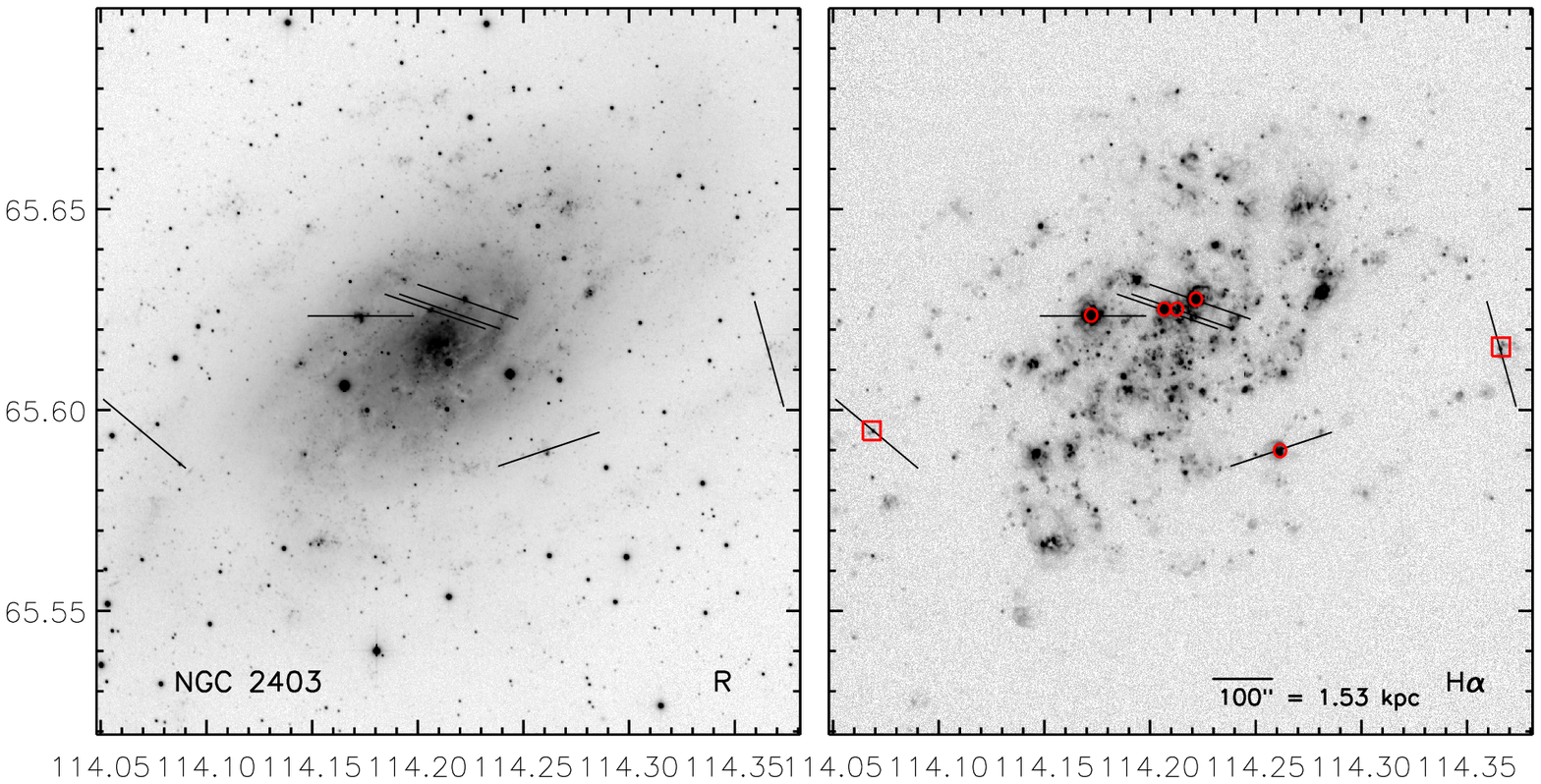}
\caption{R-band and continuum-subtracted H$\alpha$ images of NGC 2403 (van Zee et al., in prep.).
The lines represent the various slit positions observed at the MMT, 
where the central slit positions from Table~\ref{tbl3} are indicated by 
red circles for targets overlapping with the G97 sample and by 
red squares for new targets. 
See Table~\ref{tbl3} for more details.}
\label{fig2}
\end{figure}

The log information for the NGC~2403 MMT observations are tabulated in Table~\ref{tbl3}.
Figure~\ref{fig2} shows the R-band continuum and H$\alpha$ continuum-subtracted 
images for NGC~2403 (van Zee et al., in prep.), with central slit positions from Table~\ref{tbl3} indicated by 
red circles for targets overlapping with the G97 sample and by red squares
for new targets. 

%----------------------------------------------------------------------------------------------------------------------
%----------------------------------------------------------------------------------------------------------------------

% Figure 1: NGC628

%----------------------------------------------------------------------------------------------------------------------

% Table 0: Galaxy Properties
\begin{deluxetable}{lccc}
\tabletypesize{\footnotesize}
\tablewidth{0pt}
\tablecaption{Properties of NGC~628 and NGC~2403}
\startdata
{Property} 			& {NGC~628}			& {NGC~2403}  	\\
\hline													
{R.A.}				& {01:36:41.747}		& {07:36:51.400}	\\
{Dec.}				& {15:47:01.18}			& {65:36:09.20}		\\
{Type}				& {ScI}				& {SABcd}			\\
{Adopted D (Mpc)}		& {7.2$\pm$1.0}$^1$ 	& {3.16$\pm$0.07}$^6$	\\
{$m_B$ (mag)}			& {9.95}$^2$			& {8.93}$^2$		\\
{Redshift}				& {0.002192}			& {0.000445}		\\
{Inclination (degrees)}	& {5}$^3$				& {60}$^7$		\\	
{P.A. (degrees)}		& {12}$^4$			& {126}$^8$		\\
{$R_{25}$ (arcmin)}		& {5.25}$^5$			& {10.95}$^5$		\\
{$R_{25}$ (kpc)}		& {10.95}				& {10.07}			\\
\enddata
\tablecomments{Optical properties for NGC~628 and NGC~2403.
Row 1 and 2 give the RA and Dec of the optical center in units of
hours, minutes, and seconds, and decrees, arcminutes, and arcseconds respectively.
Row 5 lists redshifts taken from the NASA/IPAC Extragalactic Database.
Row 8 gives the optical radius at the $B_{25}$ mag arcsec$^{-2}$ of the system.
Row 9 gives the optical radius of the galaxy given the adopted distance.
References: (1) \citet{vandyk06}; (2) \citet{jlee11}; (3) \citet{shostak84}; (4) \citet{egusa09};
(5) \citet{kendall11}; (6) \citet{jacobs09}; (7) \citet{garnett97}; (8) \citet{fraternali02} }
\label{tbl0}
\end{deluxetable}

%----------------------------------------------------------------------------------------------------------------------

% Table 1: NGC 628 Observing Log
\begin{deluxetable}{lcccccccc}
\tabletypesize{\footnotesize}
\rotate
\tablewidth{0pt}
\tablecaption{Logs for NGC~628 Gemini Observations\label{tab:slits}}
\tablehead{
\multicolumn{9}{c}{NGC~628 GMOS Slits} }
\startdata
\CH{(1)}			& \CH{(2)}		& \CH{(3)}		& \CH{(4)}			& \CH{(5)}		& \CH{(6)}			& \CH{(7)}			& \CH{(8)}				& \CH{(9)}\\
\CH{\ion{H}{2}} 		& \CH{Alternate} & \CH{R.A.}	& \CH{Dec.}		& \CH{Slit}	& \CH{Slit PA} 		& \CH{T$_{\rm int}$}	& \CH{Offset (R.A., Dec.)} 	& \CH{$R_g$}\\
\CH{Region}		& \CH{IDs}	& \CH{(2000)} 	& \CH{(2000)} 		& \CH{Size}	& \CH{(deg)}		& \CH{(sec)}		& \CH{(arcsec)}			& \CH{(kpc)} \\
\hline 
NGC628+041-029	&	& 01:36:44.50   & 15:46:32.3	& 1.5"$\times$15"	& 0.0		& 3 $\times$ 1675	& 41.3, -28.9	& 1.77$\pm$0.25	\\ % 1-3; 	254
NGC628-034+044	&	& 01:36:39.46  	& 15:47:45.0	& 1.5"$\times$20"	& 0.0		& 3 $\times$ 1675	& -34.3, 43.8	& 1.95$\pm$0.27	\\ % 2-6;	258
NGC628+009+076	&	& 01:36:42.34  	& 15:48:17.0	& 1.5"$\times$20"	& 0.0		& 3 $\times$ 1675	& 8.9, 75.8	& 2.66$\pm$0.37	\\ % 2-5;	338
NGC628-076-029	&	& 01:36:36.71  	& 15:46:32.7	& 1.5"$\times$40"	& 0.0		& 3 $\times$ 1675	& -75.6, -28.5	& 2.83$\pm$0.39	\\ % 2-3;	65
NGC628-059+084	&	& 01:36:37.84  	& 15:48:24.7	& 1.5"$\times$20"	& 0.0		& 3 $\times$ 1675	& -58.6, 83.5	& 3.57$\pm$0.50	\\ % 2-4;	299
NGC628+082-074	&	& 01:36:47.19  	& 15:45:47.0	& 1.5"$\times$40"	& 0.0		& 3 $\times$ 1675	& 81.6, -74.2	& 3.86$\pm$0.54	\\ % 1-6;	311
NGC628+057-106	&	& 01:36:45.53  	& 15:45:15.6	& 1.5"$\times$20"	& 0.0		& 3 $\times$ 1675	& 56.7, -105.6	& 4.19$\pm$0.58	\\ % 1-5;	355
NGC628-134+069	&	& 01:36:32.82  	& 15:48:09.8	& 1.5"$\times$15"	& 0.0		& 3 $\times$ 1675	& -133.9, 68.6	& 5.27$\pm$0.73	\\ % 2-2;	180
NGC628+083-140	&	& 01:36:47.31  	& 15:44:41.6	& 1.5"$\times$20"	& 0.0		& 3 $\times$ 1675	& 83.4, -139.6	& 5.69$\pm$0.79	\\ % 1-4;	381
NGC628-044-159	&	& 01:36:38.79  	& 15:44:22.4	& 1.5"$\times$30"	& 0.0		& 3 $\times$ 1675	& -44.4, -158.8	& 5.76$\pm$0.80	\\ % 1-7;	447
NGC628-002+182	&	& 01:36:41.61  	& 15:50:03.3	& 1.5"$\times$30"	& 0.0		& 3 $\times$ 1675	& -2.1, 182.1	& 6.36$\pm$0.89	\\ % 2-8;	485
NGC628+185-052	&	& 01:36:54.11  	& 15:46:09.2	& 1.5"$\times$15"	& 0.0		& 3 $\times$ 1675	& 185.4, -52.0	& 6.75$\pm$0.94	\\ % 1-2;	220
NGC628-190+080	&	& 01:36:29.08  	& 15:48:21.4	& 1.5"$\times$15"	& 0.0		& 3 $\times$ 1675	& -190.0, 80.2	& 7.23$\pm$1.00	\\ % 2-1;	128
NGC628-090+186	&	& 01:36:35.76  	& 15:50:07.2	& 1.5"$\times$40"	& 0.0		& 3 $\times$ 1675	& -89.8, 186.0	& 7.22$\pm$1.00	\\ % 2-7 ;	434
NGC628+240+368	&	& 01:36:57.73 	& 15:47:08.9	& 1.5"$\times$50"	& 0.0		& 3 $\times$ 1675	& 239.7, 367.7	& 15.33$\pm$2.13	\\ % 1-1;	74
\newline \\
\enddata
\tablecomments{Observing logs for \ion{H}{2} regions observed in NGC~628  
using GMOS on the UT dates of 21 September 2006 and 17 November 2006.
\ion{H}{2} region ID is listed in Column 1.
Column 2 is present for listing literature IDs, but none are available.
The right ascension and declination of the individual \ion{H}{2} regions are given in units of
hours, minutes, and seconds, and decrees, arcminutes, and arcseconds respectively.
The position angle (PA) gives the rotation of the slit counter clockwise from North.
The \ion{H}{2} region distances from the center of the galaxy are listed in Column 9,
where the uncertainty in the distance has been propagated to the galactocentric radii.}
\label{tbl1}
\end{deluxetable}

%----------------------------------------------------------------------------------------------------------------------

% Table 2: NGC 628 Observing Log 2
\begin{deluxetable}{lcccccccc}
\tabletypesize{\footnotesize}
\rotate
\tablewidth{0pt}
\tablecaption{Logs for NGC~628 MMT Observations\label{tab:slits2}}
\tablehead{\multicolumn{9}{c}{NGC~628 MMT Slits} }
\startdata
\CH{(1)}			& \CH{(2)}		& \CH{(3)}		& \CH{(4)}			& \CH{(5)}		& \CH{(6)}			& \CH{(7)}			& \CH{(8)}				& \CH{(9)}\\
\CH{\ion{H}{2}} 		& \CH{Alternate} & \CH{R.A.}	& \CH{Dec.}		& \CH{Slit}	& \CH{Slit PA} 		& \CH{T$_{\rm int}$}	& \CH{Offset (R.A., Dec.)} 	& \CH{$R_g$}\\
\CH{Region}		& \CH{IDs}	& \CH{(2000)} 	& \CH{(2000)} 		& \CH{Size}	& \CH{(deg)}		& \CH{(sec)}		& \CH{(arcsec)}			& \CH{(kpc)} \\
\hline
NGC628+253+011  	& F B	& 01:37:12.49  	& 15:47:12.2 	& 1"$\times$180"	&  68.0 	& 3 $\times$ 1200 	& 253.5, 11.0	& 8.89$\pm$1.24	\\	% B-3
NGC628+267+017  	& 		& 01:37:26.40  	& 15:47:17.8 	& 1"$\times$180"	&  68.0 	& 3 $\times$ 1200 	& 267.4, 16.6	& 9.38$\pm$1.31	\\	% B-2
NGC628+277+020   & 		& 01:37:35.67  	& 15:47:21.6 	& 1"$\times$180"	&  68.0 	& 3 $\times$ 1200 	& 276.7, 20.4	& 9.72$\pm$1.35	\\	% B-1
NGC628+295-016	& vZ 6	& 01:37:01.39  	& 15:46:45.5 	& 1"$\times$180"	&  30.0 	& 3 $\times$ 1800 	& 294.6, -15.7	& 10.34$\pm$1.44	\\	% vZ6
NGC628-277+241	& F C	& 01:31:45.97  	& 15:55:01.8 	& 1"$\times$180"	&  72.0 	& 3 $\times$ 1800 	& -277.3, 240.7	& 12.85$\pm$1.79	\\	% C-4
NGC628-288+240	& 		& 01:31:35.31  	& 15:55:01.0 	& 1"$\times$180"	&  72.0 	& 3 $\times$ 1800 	& -288.0, 239.9	& 13.12$\pm$1.82	\\	% C-2
NGC628+185+356	& F E	& 01:39:59.43  	& 15:58:53.1 	& 1"$\times$180"	&  -26.0 	& 3 $\times$ 1800 	& 185.3, 355.8	& 14.01$\pm$1.95	\\	% E-6
NGC628+186+425	& 		& 01:39:59.88  	& 16:00:02.1 	& 1"$\times$180"	&  -26.0 	& 3 $\times$ 1800 	& 185.7, 424.8	& 16.19$\pm$2.25	\\	% E-4
NGC628+503+208	& F F	 	& 01:37:15.26  	& 15:50:29.1 	& 1"$\times$180"	&  -35.0 	& 3 $\times$ 1800 	& 502.7, 207.9	& 19.04$\pm$2.65	\\	% F-1
\enddata
\tablecomments{Observing logs for \ion{H}{2} regions observed in NGC~628 at the MMT 
on the UT dates of 31 October 2008.
\ion{H}{2} region IDs are listed in Column 1.
Other literature IDs are given in Column 2:
\citet[][F]{ferguson98} and \citet[][vZ]{vanzee98b}.
The right ascension and declination of the individual \ion{H}{2} regions are given in units of
hours, minutes, and seconds, and decrees, arcminutes, and arcseconds respectively.
The position angle (PA) gives the rotation of the slit counter clockwise from North.
The \ion{H}{2} region distances from the center of the galaxy are listed in Column 9,
where the uncertainty in the distance has been propagated to the galactocentric radii.}
\label{tbl2}
\end{deluxetable}

%----------------------------------------------------------------------------------------------------------------------

% Figure 2: NGC 2403

%----------------------------------------------------------------------------------------------------------------------

% Table 3: NGC 2403 Observing Log
\begin{deluxetable}{lcccccccc}
\tabletypesize{\footnotesize}
\rotate
\tablewidth{0pt}
\tablecaption{Logs for NGC~2403 MMT Observations\label{tab:slits3}}
\tablehead{
\multicolumn{9}{c}{NGC~2403 MMT Slits} }
\startdata
\hline
\CH{(1)}			& \CH{(2)}		& \CH{(3)}		& \CH{(4)}			& \CH{(5)}		& \CH{(6)}			& \CH{(7)}			& \CH{(8)}				& \CH{(9)}\\
\CH{\ion{H}{2}} 		& \CH{Alternate} & \CH{R.A.}	& \CH{Dec.}		& \CH{Slit}	& \CH{Slit PA} 		& \CH{T$_{\rm int}$}	& \CH{Offset (R.A., Dec.)} 	& \CH{$R_g$}\\
\CH{Region}		& \CH{IDs}	& \CH{(2000)} 	& \CH{(2000)} 		& \CH{Size}	& \CH{(deg)}		& \CH{(sec)}		& \CH{(arcsec)}			& \CH{(kpc)} \\
\hline 
NGC2403-007+036	& VS 35, HK 313 	& 07:36:50.3  	& 65:36:45	& 1"$\times$180"	& -71.5	& 6 $\times$ 900	& -7, 36		& 0.87$\pm$0.02	\\ 	% 35
NGC2403-030+045	& VS 24, HK 361	& 07:36:46.6   	& 65:36:54	& 1"$\times$180"	& -71.5	& 6 $\times$ 900	& -30, 45		& 0.97$\pm$0.02	\\	% 24
NGC2403+013+031	& VS 38, HK 270	& 07:36:53.5  	& 65:36:40	& 1"$\times$180"	& -71.5	& 6 $\times$ 900	& 13, 31		& 1.01$\pm$0.02	\\	% 38
NGC2403+104+024	& VS 44, HK 128	& 07:37:08.2  	& 65:36:33	& 1"$\times$180"	& -90.0	& 11 $\times$ 300	& 104, 24		& 2.69$\pm$0.06	\\	% 44
NGC2403-133-146	& VS 9			& 07:36:29.9  	& 65:33:43	& 1"$\times$180"	& -108.0	& 6 $\times$ 300	& -133, -146	& 6.02$\pm$0.13	\\	% 9
NGC2403+376-106	& HK 376			& 07:37:52.1  	& 65:34:23	& 1"$\times$180"	& -231.0	& 3 $\times$ 900	& 376, -106	& 6.98$\pm$0.16 	\\	% 376
NGC2403-423-010	& HK 423			& 07:35:43.1  	& 65:35:59	& 1"$\times$180"	& -231.0	& 5 $\times$ 1800	&-423, -10	& 9.40$\pm$0.21	\\	% 423
\enddata
\tablecomments{Observing logs for \ion{H}{2} regions observed in NGC 2403 at the MMT 
on the UT dates of 31 October 2008 and 1-2 November 2008 and using GMOS on the 
UT dates of 21 September 2006 and 17 November 2006,
and observing log for \ion{H}{2} regions observed in NGC 2403 at the MMT on the UT
dates of 2006 February 1-4.
\ion{H}{2} region IDs are listed in Column 1.
Other literature IDs are given in Column 2:
\citet[][VS]{veron65} and \citet[][HK]{hodge83}.
The right ascension and declination of the individual \ion{H}{2} regions are given in units of
hours, minutes, and seconds, and decrees, arcminutes, and arcseconds respectively.
The position angle (PA) gives the rotation of the slit counter clockwise from North.
The \ion{H}{2} region distances from the center of the galaxy are listed in Column 9,
where the uncertainty in the distance has been propagated to the galactocentric radii.}
\label{tbl3}
\end{deluxetable}

%----------------------------------------------------------------------------------------------------------------------
%----------------------------------------------------------------------------------------------------------------------
\subsection{Spectra Reduction}\label{sec:reduct}

GMOS spectra were reduced and extracted using the {\tt GMOS} package of 
IRAF\footnote[2]{IRAF is distributed by the National Optical Astronomical Observatories.}. 
Reduction of spectra included bias subtraction and flat fielding 
based on observations of the quartz halogen lamp on the GCAL unit.  
A spectral trace of a bright continuum source was used to define the 
trace for all slits in each field. 
The shape of this trace was consistent in all exposures of a given field.  
As the slits were long compared to the spatial extent of the individual 
\ion{H}{2} regions, the sky background was removed from the 
two-dimensional spectra via the subtraction of local measurements 
adjacent to the \ion{H}{2} region.  
This local sky subtraction does have the potential to over-subtract strong 
lines relative to weaker emission lines, as the temperature of the diffuse 
ISM may be different from the temperature within the \ion{H}{2} regions. 
However, the contributions to measured emission lines from the diffuse 
ISM is deemed to be negligible as all slits showed comparable 
backgrounds to the two slits that were both long and situated at the 
edge of NGC~628's disk. 
While red and blue spectra were not observed simultaneously, 
we note that the continuum is well matched where spectra overlap, 
indicating stable sky conditions and matched extraction apertures.

\begin{figure}[H]
\epsscale{1.0}
\figurenum{3}
\plotone{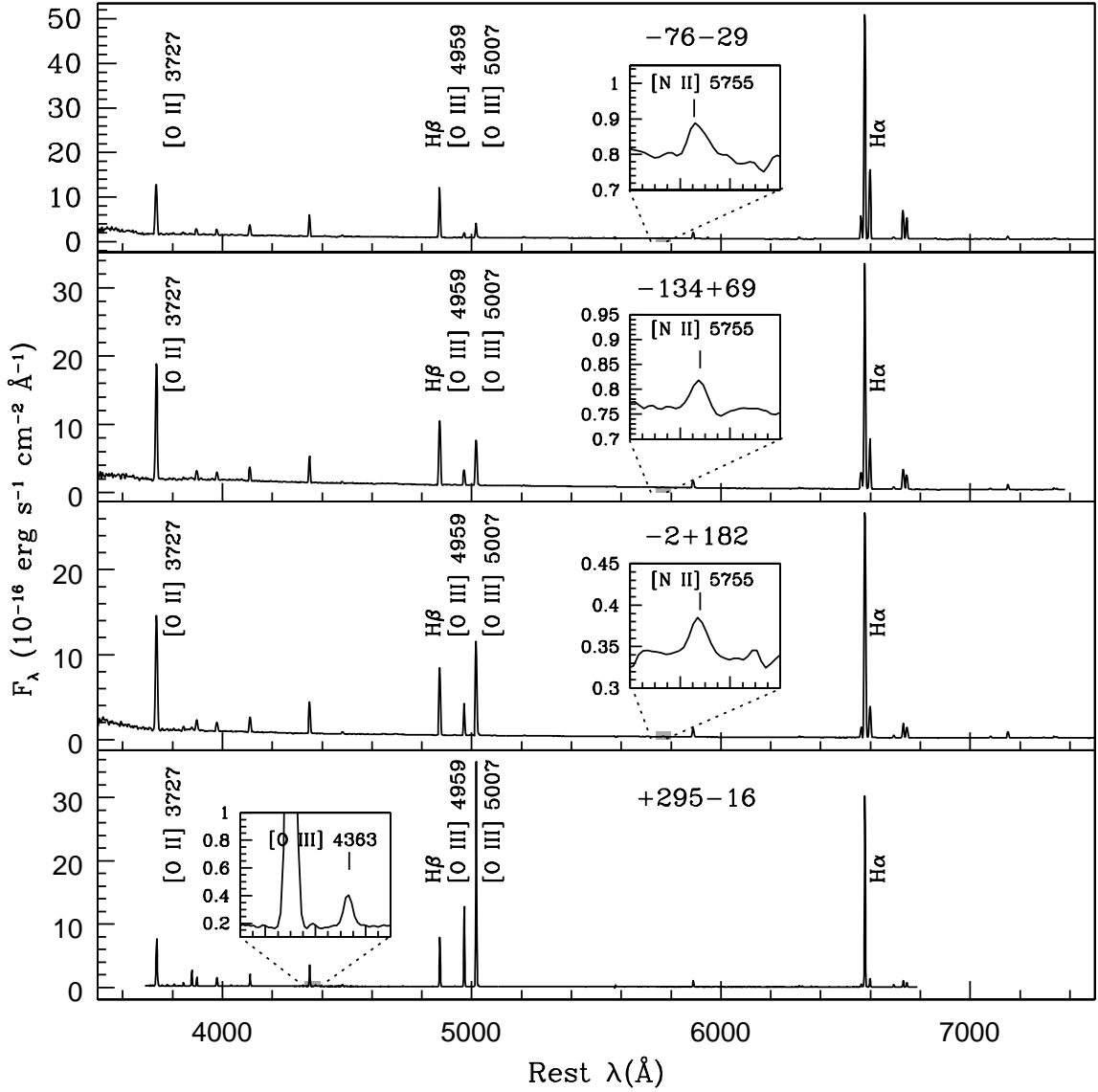}
\caption{Sample MMT and GMOS spectra of the \ion{H}{2} regions observed across the 
face of NGC 628 with auroral line detections at a strength of 4$\sigma$ or greater.
The top panel is at a galactocentric radius of 2.83 kpc and has the smallest [\ion{O}{3}]/[\ion{O}{2}]
ratio, indicative of low excitation. Each subsequent spectra is at larger radii and has higher excitation.
The [\ion{O}{3}] $\lambda4363$ and [\ion{N}{2}] $\lambda5755$ auroral lines are 
expanded to highlight the quality of these spectra.}
\label{fig3}
\end{figure}

The NGC 628 MMT observations were processed using ISPEC2D 
\citep{moustakas06}, a long-slit spectroscopy data reduction package 
written in IDL, as described in \cite{berg11} and \cite{berg12}. 
Figure~\ref{fig3} shows a sample of the resulting one-dimensional spectra 
extracted for \ion{H}{2} regions in NGC~628 which had significant 
[\ion{O}{3}] $\lambda$4363 or [\ion{N}{2}] $\lambda$5755 detections. 
The inset windows display a narrower spectral range to 
emphasize the auroral line strengths.
The NGC 2403 MMT observations were processed following 
\citet{garnett97}, with no significant differences. 

%----------------------------------------------------------------------------------------------------------------------
%----------------------------------------------------------------------------------------------------------------------

% Figure 3: Sample Spectra

%----------------------------------------------------------------------------------------------------------------------
%----------------------------------------------------------------------------------------------------------------------

%----------------------------------------------------------------------------------------------------------------------

\section{NEBULAR ABUNDANCE ANALYSIS}\label{sec:abund}

%----------------------------------------------------------------------------------------------------------------------

\subsection{Emission Line Measurements}\label{sec:iraf}
Emission line strengths were measured using standard methods available within IRAF.
Specifically, the {\tt SPLOT} routine was used to analyze the extracted 
one-dimensional spectra and fit Gaussian profiles 
to emission lines to determine their integrated fluxes. 
Special attention was paid to the Balmer lines, which can be 
located in troughs of significant underlying stellar absorption. 
In the cases where Balmer absorption was clearly visible, the bluer Balmer lines 
(H$\delta$ and H$\gamma$) were fit simultaneously with multiple components where the absorption was 
fit by a broad, negative Lorentzian profile and the emission was fit by a narrow, positive Gaussian profile. 
The negative absorption flux is measure to be roughly consistent for all Balmer lines of a given spectrum,
and so for stronger emission lines the absorption component is too weak to measure.
This was the case for H$\beta$ and H$\alpha$ in all of our spectra;
the absorption component was negligible and not measured.
Additional constraints were placed on the faint lines, including
the [\ion{O}{3}] $\lambda4363$ and [\ion{N}{2}] $\lambda5755$ auroral
lines, such that their FWHMs match the neighboring strong line fits.

\citet{skillman94} give a strict accounting of the standard errors ($\delta L$)
associated with a given line strength ($L$):
\begin{equation}
	\frac{\delta L}{L} = \frac{ \{C_1 + C_2 + (n_o/\sqrt{n_s})S + nAN^2 + [2.3f(\lambda)\delta C(\mbox{H}\beta)L]^2 + [0.01L]^2 + [\delta FL]^2\}^{1/2}}{L},
\end{equation}
where $S$ is the sky background in a single row and $n_o$ and $n_s$ are 
the number of rows summed over for the object and the sky respectively.
Due to the quality of the detectors used the read out noise term ($nAN^2$) 
is negligible compared to the other dominating terms.
For weak lines, the uncertainty is dominated by error from the continuum subtraction,
meaning that the $C_2$ (counts in the continuum) term is the dominating fraction of
the $C_1$ (total counts in line, continuum, and sky) term.
For the lines with flux measurements much stronger than the rms noise of the continuum, 
(usually the H$\alpha$ lines and often the [\ion{O}{3}] $\lambda\lambda$4959,5007 doublet) 
the error is dominated by flux calibration error term ($[\delta FL]^2$) and the 
reddening error term ($[2.3f(\lambda)\delta C(H\beta)L]^2$). 
Thus, in the case of our spectra, the errors of the flux measurements were approximated using 
\begin{equation}
	\delta F_{\lambda} \approx \sqrt{ {(2\times \sqrt{n_p}\times rms)}^2 + {(0.02\times F_{\lambda})}^2 } ,	\label{eq:uncertainty}
\end{equation}
where $n_p$ is the number of pixels spanning the 
Gaussian profile fit to the narrow emission lines. 
The root mean squared ($rms$) noise in the continuum was taken to be the 
average of the rms on each side of an emission line. 
Then, the first ($rms$) term determines the approximate uncertainty for the weak lines,
and the second ($F_{\lambda}$) term dominates the uncertainty for the strong lines.
In the latter case, a minimum uncertainty of 2\% was assumed, which is the typical
error in fitting flux calibrated points to standard \citet{oke90} stars.

11 \ion{H}{2} regions in our NGC~628 sample were measured 
to have [\ion{O}{3}] $\lambda$4363 and/or [\ion{N}{2}] $\lambda$5755 
line strengths $> 4\sigma$; three additional objects had auroral line 
strengths $> 3\sigma$.
In NGC~2403, 7 of the \ion{H}{2} regions auroral line 
detections had strengths of $4\sigma$ or greater.
Note that the [\ion{O}{3}] $\lambda4363$ line is often difficult to 
detect due to strong mercury line contamination ($\lambda4358$) 
at some observatories, however, NGC~628 has a large enough 
redshift (z$\sim0.0022$) to clearly distinguish the lines. 
The redshift for NGC~2403 is lower (z$\sim0.0004$), but this concern 
is negated by the fact that six of the seven \ion{H}{2} regions in our sample 
also have a temperature measurement from [\ion{N}{2}] $\lambda5755$.
For all of the objects in the present samples, flux line strengths and 
corresponding errors are listed in Tables~\ref{tbl4}-\ref{tbl6}. 
We concentrate the rest of our analysis on the objects for which direct electron 
temperature and chemical abundance determinations can be made.

%----------------------------------------------------------------------------------------------------------------------

\subsection{Reddening Corrections}\label{sec:redcor}
The relative intensities of the Balmer lines are nearly independent of both 
density and temperature, so they can be used to solve for the reddening. 
The spectra were de-reddened using the reddening law of \citet{cardelli89}, 
parametrized by $A_{V}=3.1\ E(B-V)$, where the extinction, $A_{1}(\lambda)$ 
was calculated using the York Extinction Solver 
\citep{mccall04}\footnote{http://www1.cadc-ccda.hia-iha.nrc-cnrc.gc.ca/community/YorkExtinctionSolver/}. 
With these values, the reddening, $E(B-V)$, can be derived using
\begin{equation}
	\log{\frac{I(H\alpha)}{I(H\beta)}\ } = \log{\frac{F(H\alpha)}{F(H\beta)}\ } + 0.4\ E(B-V)\ [A_{1}(H\alpha)-A_{1}(H\beta)],
	\label{eq:dered} 
\end{equation}
where $F$(H$\alpha)/F$(H$\beta$) is the observed flux ratio and 
$I$(H$\alpha)/I$(H$\beta$) is the de-reddened line intensity ratio 
using case B from \citet{hummer87}, assuming an electron temperature 
calculated from the [\ion{O}{3}] line ratio and n$_{e}=10^{2}$ cm$^{-3}$.
For our NGC 628 sample, the electron temperature covers a range of 6,300 K to 14,100 
K in the high ionization zone and from 7,400 K to 13,300 K in the low ionization zone.
NGC~2403 exhibits a smaller electron temperature range from 7,700 K to 11,300 K 
in the high ionization zone and from 8,300 K to 10,900 K in the low ionization zone 
(see discussion of electron temperatures in Section~3.3).
Following \citet{lee04}, the reddening value can be converted to the 
logarithmic extinction at $H\beta$ as
\begin{equation}
	c(\mbox{H}\beta) = 1.43\ E(B-V).
	\label{eq:cHbeta}
\end{equation}
We evaluate our reddening corrections by using a chi-squared minimization
technique in comparing calculated to theoretical values to ensure 
the best combined solution for the H$\delta$/H$\beta$, H$\gamma$/H$\beta$,
and H$\alpha$/H$\beta$ ratios \citep[see discussion in][]{olive01}.
For those cases where underlying absorption was accounted for in measuring the
individual emission lines, we fix underlying absorption to be zero in the minimization;
otherwise reddening and underlying absorption were solved for simultaneously.
For NGC~628 we find a range in $A_V$ of 0.13 to 1.56 and a slightly 
smaller range for NGC~2403 of 0.04 to 1.26.
The results are tabulated in Tables~\ref{tbl4}-\ref{tbl6}.

%----------------------------------------------------------------------------------------------------------------------
%----------------------------------------------------------------------------------------------------------------------

\subsection{Electron Temperature and Density Determinations}\label{sec:temden}
For \ion{H}{2} regions, accurate direct oxygen abundance determinations 
require reliable electron temperature measurements.
This is typically done by observing a temperature sensitive auroral line.
We have taken advantage of two such lines: the [\ion{O}{3}] $\lambda$4363
and [\ion{N}{2}] $\lambda$5755 auroral lines.
To establish a low-uncertainty electron temperature estimate, we define 
a strong auroral line measurement using a 4$\sigma$ criterion.
In NGC~628, we measured [\ion{O}{3}] $\lambda$4363 at a strength of 
4$\sigma$ or greater in five \ion{H}{2} regions and measured [\ion{N}{2}] 
$\lambda$5755 at a strength of 4$\sigma$ or greater in ten \ion{H}{2} regions.
Three of these \ion{H}{2} regions had both strong [\ion{O}{3}] $\lambda$4363
and [\ion{N}{2}] $\lambda$5755 such that we were able to measure 11 total
\ion{H}{2} regions with direct abundances from at least one strong auroral line. 
We extend our NGC~628 sample to 14 by including 3 more \ion{H}{2} regions
that have a 3$\sigma$ detections of the auroral line.

In NGC~2403, we measured [\ion{O}{3}] $\lambda$4363 at a strength of 
4$\sigma$ or greater in five \ion{H}{2} regions and measured [\ion{N}{2}] 
$\lambda$5755 at a strength of 4$\sigma$ or greater in six \ion{H}{2} regions
such that all seven \ion{H}{2} regions have strong auroral line measurements.
For the \ion{H}{2} regions in NGC~628 and NGC~2403 with strong auroral lines, 
we then determined electron temperatures from the associated temperature 
sensitive ``auroral" to ``nebular" ratio of collisionally excited lines.

An \ion{H}{2} region can be modeled by two separate 
volumes, one of low ionization and one of high ionization.
For the high ionization zone, we used the [\ion{O}{3}] 
I($\lambda\lambda$4959,5007)/I($\lambda$4363) ratio to 
derive a temperature using the IRAF task {\tt TEMDEN}. 
This task computes the electron temperature of the ionized 
nebular gas within the 5-level atom approximation.
The O$^{+}$ (low ionization) zone electron temperature can be 
related to the O$^{++}$ (high ionization) zone electron temperature 
\citep[e.g.,][]{campbell86,pagel92}.
We used the relation between $t_{2}$ ($T_{e}$(O$^{+}$)) and 
$t_{3}$ ($T_{e}$(O$^{++}$)) proposed by \citet{pagel92}, based 
on the photoionization modeling of \citet{stasinska90} to determine 
the low ionization zone temperature:
\begin{equation}
	t_e(\mbox{\ion{O}{2}})^{-1} = 0.5\times[{t_e(\mbox{\ion{O}{3}})}^{-1} + 0.8],
	\label{eqn:OII}
\end{equation}
where $t_{e} = T_{e}/10^4$ K.

For six of our spectra in NGC 628 and one spectrum in NGC 2403 
where a strong [\ion{O}{3}]$\lambda$4363 was not  measured, 
we were able to determine a temperature using the [\ion{N}{2}] 
I($\lambda\lambda$6548,6584)/I($\lambda$5755) ratio,
where we assume $T_e$(\ion{N}{2}) = $T_e$(\ion{O}{2}).
The high ionization zone temperature can then be determined 
using Equation~\ref{eqn:OII} in reverse.
Both the [\ion{O}{3}] and [\ion{N}{2}] auroral lines were measured 
in four \ion{H}{2} regions in NGC~628 and five \ion{H}{2} regions 
in NGC~2403, allowing us to compare electron temperature estimates 
for the high ionization zone from two different diagnostics. 
On average, [\ion{O}{3}] predicts a higher electron temperature by 
1,100 K in NGC~628 and by 700 K in NGC~2403.
While we prioritize the [\ion{O}{3}] temperature estimate in this paper,
further studies are needed to determine for what cases the [\ion{N}{2}] 
temperature diagnostic is more reliable, and where the differences originate.
Note that [\ion{O}{2}] $\lambda\lambda7320,7330$ was measured in multiple
\ion{H}{2} regions, however, \citet{zurita12} have shown that the error in using
[\ion{O}{2}] is much larger than in using [\ion{N}{2}] to derive a temperature for
the low ionization zone. 
In the five \ion{H}{2} regions in NGC~628 and four \ion{H}{2} regions 
in NGC~2403 with temperature estimates from both [\ion{O}{2}] and 
[\ion{N}{2}], we find that the $T_{e}$(\ion{O}{2}) estimates are 900 K 
and 600 K larger on average than those from [\ion{N}{2}]. 
This may be due in large part to the fact that the [\ion{O}{2}] 
$\lambda\lambda7320,7330$ doublet is located in a spectral 
region containing strong OH airglow emission lines.
Additionally, differences in the electron temperature determinations
may originate from errors in the standard \ion{H}{2} region physics assumed.
\citet{peimbert67} introduced the temperature inhomogeneity parameter,
$t^2$, to be used as a statistical correction to chemical abundances. 
In order to account for temperature and ionization structures which
differ from standard assumptions, \citet{pena-guerrero12} used $t^2$ 
corrections to re-calibrate the strong-line method of \citet{pagel85}.
In a different attempt, \citet{nicholls12} suggest that the electrons in a \ion{H}{2} 
region may depart from a Maxwell-Boltzmann equilibrium energy distribution,
and instead suggest the adoption of a ``$\kappa$-distribution". 
However, further work is needed to ascertain the most effective solution.

Additionally, we calculate a separate $T_{e}$(\ion{S}{3}) because the S$^{++}$ 
ionization zone lies partially in both the O$^+$ and the O$^{++}$ zones:
\begin{equation}
	t_{e}(\mbox{\ion{S}{3}}) = 0.83\times t_{e}(\mbox{\ion{O}{3}}) + 0.17.
\end{equation}
The above temperatures are tabulated in Tables~\ref{tbl7}-\ref{tbl9}. 
[\ion{S}{2}] $\lambda\lambda$6717,6731 was used to determine the 
electron densities, all of which are consistent with the low density limit.
For all abundance calculations we assume $n_e = 10^2$ cm$^{-3}$ 
(which is consistent with the 1$\sigma$ upper bounds and produces 
identical results for all lower values of $n_e$).

\subsection{Ionic and Total Abundances}

Ionic abundances were calculated with:
\begin{equation}
	{\frac{N(X^{i})}{N(H^{+})}\ } = {\frac{I_{\lambda(i)}}{I_{H\beta}}\ } {\frac{j_{H\beta}}{j_{\lambda(i)}}\ }.
	\label{eq:Nfrac}
\end{equation}
The emissivity coefficients, which are functions of both temperature 
and density, were determined using the {\tt IONIC} routine in IRAF 
with atomic data updated as reported in \citet{bresolin09a}.
Note that \citet{nicholls13} discuss the the factors involved in obtaining accurate 
temperatures from collisionally excited lines, showing that significant errors arise 
when using old collision strength data, and thus using the updated
atomic data is crucial.
This routine applies the 5-level atom approximation, assuming 
an electron density of n$_{e}=10^2$ cm$^{-3}$ and the 
appropriate ionization zone electron temperature.

Total oxygen abundances (O/H) are calculated from the simple 
sum of O$^{+}$/H$^{+}$ and O$^{++}$/H$^{+}$.
The other abundance determinations require ionization correction 
factors (ICF) to account for unobserved ionic species. 
For nitrogen, we employ the common assumption that N/O 
= N$^{+}$/O$^{+}$ \citep{peimbert67}.
This allows us to directly compare our results with other studies in the literature.
\cite{nava06} have investigated the validity of this assumption.  
They concluded that it could be improved upon with modern 
photoionization models, but was valid at the precision of about 10\%.

Collisionally excited emission lines of sulfur, neon, and 
argon were also observed in many of our spectra.
For Ne we use a fairly straightforward ICF: ICF(Ne) = 
(O$^{+}$ + O$^{++}$)/O$^{++}$ \citep{crockett06}.
S and Ar present more complicated situations as S$^{++}$ and 
Ar$^{++}$ span both the O$^{+}$ and the O$^{++}$ zones.
\citet{thuan95} have determined the analytic ICF approximations 
for both S and Ar using the model calculations of 
photoionized \ion{H}{2} regions by \citet{stasinska90}.
We employ these ICFs from \citet{thuan95} to correct for 
the unobserved S$^{+3}$,  Ar$^{+2}$, and Ar$^{+4}$ states.

Ionic and total abundances are listed in Tables~\ref{tbl7}-\ref{tbl9}.
We report  direct oxygen abundances for 14 
\ion{H}{2} regions in NGC~628 (11 of which meet our 4$\sigma$ criterion) 
and new direct oxygen abundances for seven NGC~2403 \ion{H}{2} regions. 

%----------------------------------------------------------------------------------------------------------------------

\section{THE ABUNDANCE GRADIENTS IN NGC~628 AND NGC~2403}

\subsection{Oxygen}
\label{sec:O}

To analyze the oxygen abundance gradients for NGC~628 and NGC~2403
we have plotted the $3\sigma$ (open circles) and $4\sigma$ (filled symbols)
direct abundance detections versus galactocentric radius in 
Figures~\ref{fig4} and \ref{fig5} respectively.
We determine the most likely linear fits to the data using the FITEXY routine in IDL.
This routine uses a least-squares minimization in one-dimension where both 
independent and dependent variables have associated errors.
For the present abundance gradients, the uncertainties in oxygen abundance 
and galactocentric radius, propagated from the uncertainty of the distance to the galaxy,
have been used as weights.

\subsubsection{NGC 628}

\begin{figure}[H]
\figurenum{4}
\epsscale{0.6}
\plotone{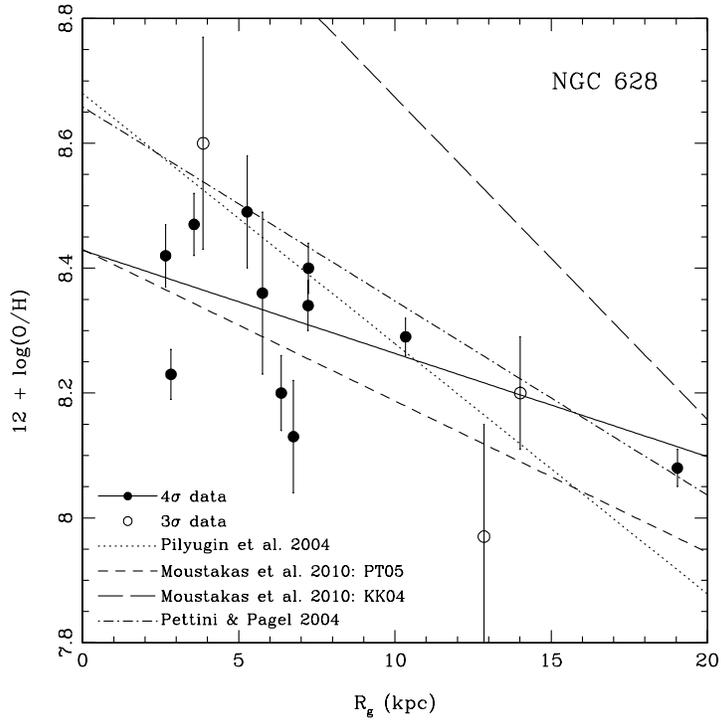}
\caption{Direct oxygen abundances from Gemini and the MMT 
are plotted versus galactocentric radius for NGC 628. 
Solid points have auroral line detections at a strength at 4$\sigma$ or 
greater and the open points have detections at strengths of 3$\sigma$.
The solid line depicts the least-squares best fit to the 4$\sigma$ data.
For comparison, the strong-line oxygen abundance gradient determined by 
\citet{pilyugin04} using the P-method \citep{pilyugin01} is plotted as a dotted line and 
the gradients from \citet{moustakas10} using the \citet{pilyugin05} and \citet{kobulnicky04} 
P-methods are plotted as short-dashed and long-dashed lines respectively.
Additionally, we used the N2 calibration of \citet{pp04} to determined the strong-line 
abundance gradient for the present observations and plotted the fit as a dotted-dashed line.
Note the difference between the direct and strong-line relationships, which is especially significant 
considering no other solely direct abundance studies have been published for NGC 628.}
\label{fig4}
\end{figure}

In Figure~\ref{fig4} we plot direct oxygen abundance 
versus galactocentric radius for NGC~628.
The filled points represent all of the targets with direct abundances 
measured at a strength of 4$\sigma$ or greater and the open points 
are three additional \ion{H}{2} regions with auroral line detections 
at a strength of 3$\sigma$.
Our sample is fairly well dispersed over the range of $R_g\sim2-23$
 kpc, containing 10 \ion{H}{2} regions within the isophotal radius 
($R_{25}\sim10.25$ kpc) and four extending beyond $R_{25}$.
As expected from studies of spiral galaxies, the innermost \ion{H}{2} 
regions tend to have higher oxygen abundances and the outermost 
\ion{H}{2} regions have comparatively low oxygen abundances.

The best fit to characterize the gradient of the 11 objects in the current 
4$\sigma$ sample with direct oxygen abundance measurements is given by:
\begin{equation}
	12 + \log(\mbox{O/H}) = (8.43\pm0.03) + (-0.017\pm0.002)\times{R_g}  \mbox{ (dex/kpc)},
\end{equation}
with a dispersion in log(O/H) of $\sigma$ = 0.10 dex.
Additionally, we fit the 3$\sigma$ sample and found no difference in the 
resulting fit within the significant digits that we quote here.
While the addition of the $3\sigma$ does not alter the fit, we choose to 
fit only the $4\sigma$ data in order to protect against false detections
of the weak auroral lines.

The gas-phase oxygen abundance of NGC~628 has been
previously studied using empirical calibrations
\citep[e.g.,][]{talent83,mccall85,zaritsky94,ferguson98,vanzee98b,
bresolin99,castellanos02,pilyugin04,moustakas10,gusev12,cedres12},
and integral field spectroscopy \citep[e.g.,][]{rosales-ortega11}.
For comparison, we have plotted the strong-line relationships of \citet{kobulnicky04}
and \citet{pilyugin05} for NGC 628 as compiled by \citet{moustakas10} from literature values.
Both methods are based on the metallicity-sensitive $R_{23}$ parameter \citep{pagel79},
with an additional excitation parameter that corrects for ionization (referred to as the ``P-method").
Using optical spectroscopy of 34 \ion{H}{2} regions, \citet{moustakas10} found 
\begin{equation}
	12 + \log(\mbox{O/H}) = 9.19 - 0.052\times{R_g}  \mbox{ (dex/kpc)},
\end{equation}
for the \cite{kobulnicky04} relationship and
\begin{equation}
	12 + \log(\mbox{O/H}) = 8.43 - 0.024\times{R_g}  \mbox{ (dex/kpc)},
\end{equation}
for the \cite{pilyugin05} relationship, based on 33 \ion{H}{2} regions.

\citet{pilyugin04} found a significantly steeper gradient in NGC 628
using an earlier version of the P-method \citep{pilyugin01}:
\begin{equation}
	12 + \log(\mbox{O/H}) = 8.68 - 0.040\times{R_g}  \mbox{ (dex/kpc)}.
\end{equation}
However, \citet{bresolin12} used observations of NGC 1512 and NGC 3621 
to show that, of the strong line methods in wide use, the N2 calibration 
of \citet{pp04} best matches the direct abundances.
We have applied this method to our observations and plotted the least-
squares fit to the results as a dotted-dashed line in Figure~\ref{fig4}.
Interestingly, this fit lies close to the direct abundances, but has a 
steeper slope and an approximately 0.2 dex high intercept at the center.
\citet{rosales-ortega11} used a variety of methods with integral field 
spectroscopy to determine the radial oxygen abundance gradient for NGC 628.
Using their ff-$T_e$ method, they calculated a linear fit of
\begin{equation}
	12 + \log(\mbox{O/H}) = (8.70\pm0.01) + (-0.026\pm0.001)\times{R_g}  \mbox{ (dex/kpc)}.
\end{equation}
\citet{rosales-ortega11} also examined the gradient using three 
other strong-line calibrators, finding that the slope of the gradient 
varies significantly among different calibrators. 
They conclude that this may be due to the potential for empirical 
indices based on the [\ion{N}{2}] emission lines to overestimate 
oxygen abundance at high N/O ratios and vice versa 
\citep[see e.g.,][]{perez-montero09}.
This is easily seen in our Figure~\ref{fig6} as the \cite{pilyugin05} 
and \cite{moustakas10} relationships extend to much higher oxygen 
abundances with increasing N/O .

%----------------------------------------------------------------------------------------------------------------------
%----------------------------------------------------------------------------------------------------------------------

% Figure 4: NGC 628 Oxygen Gradient

%----------------------------------------------------------------------------------------------------------------------
%----------------------------------------------------------------------------------------------------------------------

\subsubsection{NGC 2403}

The present sample includes three \ion{H}{2} regions with reported 
T(\ion{O}{3}) in common with G97: VS~24, VS~44, and VS~9.  
A comparison of the derived electron temperatures and oxygen 
abundances  shows general, but not perfect agreement.  
Specifically, for VS~24, G97 derived an \ion{O}{3} temperature of 7,600 $\pm$ 400 K 
and an oxygen abundance of 8.41 $\pm$ 0.09 compared with a temperature of 
7,700 $\pm$ 700 K and an oxygen abundance of 8.45 $\pm$ 0.11derived here.
For VS~44, G97 derived an \ion{O}{3} temperature of 8,300 $\pm$ 400 and an 
oxygen abundance of 8.49 $\pm$ 0.09 compared with a temperature of 
8,700 $\pm$ 200 and an oxygen abundance of 8.40 $\pm$ 0.04 derived here.
For VS~9, G97 derived an \ion{O}{3} temperature of 11,700 $\pm$ 400 and an 
oxygen abundance of 8.10 $\pm$ 0.03 compared with a temperature of 
11,100 $\pm$ 200 and an oxygen abundance of 8.28 $\pm$ 0.04 derived here.
Thus, all of the above measurements agree within 1$\sigma$, 
except for the oxygen abundance of VS~9.  
As it turns out, G97 did not measure a blue spectrum for VS~9, but, instead,
combined the blue spectrum from \citet{mccall85} with a red spectrum from the MMT.
Thus, from the two spectra in common, it appears that the potential non-linearities
in the detector are not a concern and that the G97 observations can be compared directly.
Given the above, we derive the oxygen abundance gradient both with (1) only 
the newly obtained observations and (2) from combining the new observations 
with the previous observations obtained by G97.

In Figure~\ref{fig5} we plot direct oxygen abundance 
versus galactocentric radius for NGC~2403.
Our observations are depicted by solid symbols, 
the data in common with G97 are given by filled circles 
and new regions by filled squares.
Additional direct observations made by G97 are plotted as open circles.
Combining these data, there are 11 points displayed for the inner 
galaxy extending out to $\sim10$ kpc.
Again we see decreasing oxygen abundance with increasing 
radius, but with a slightly steeper gradient than that of NGC 628.

The weighted least-squares fit (dotted dahsed line) to the seven new observations 
(filled symbols) presented here can be expressed as:
\begin{equation}
	12 + \log(\mbox{O/H}) = (8.46\pm0.04) + (-0.027\pm0.008)\times{R_g}  \mbox{ (dex/kpc)}, \label{O2403_Us}
\end{equation}
with a dispersion in log(O/H) of $\sigma$ = 0.02 dex.
In comparison, the best weighted fit (solid line) to characterize the 
gradient of the 11 points, both new and old, in NGC 2403 with direct 
oxygen abundance measurements results in:
\begin{equation}
	12 + \log(\mbox{O/H}) = (8.48\pm0.04) + (-0.032\pm0.007)\times{R_g}  \mbox{ (dex/kpc)}, \label{O2403_Tot}
\end{equation}
with a dispersion in log(O/H) of $\sigma$ = 0.07 dex.
This result is equivalent to the relationship derived for just the 
new observations, so we carry forward with the addition of the 
G97 observations to increment our sample. 

For NGC~2403, several spectroscopic studies exist where 
abundances were determined using empirical calibrations 
\citep[e.g.,][]{mccall85,fierro86,vanzee98b,garnett99,bresolin99}, 
as well as a single study using direct abundances \citep{garnett97}. 
The dotted line in Figure~\ref{fig5} is the fit of \citet{pilyugin04} 
using the strong-line P-method \citep{pilyugin01}:
\begin{equation}
	12 + \log(\mbox{O/H}) = 8.52 - 0.033\times{R_g}  \mbox{ (dex/kpc)}. 
\end{equation}
The dashed line is the fit of \citet{moustakas10} using the 
strong line method from \citet{pilyugin05}:
\begin{equation}
	12 + \log({\mbox{O/H}}) = 8.42 - 0.032\times{R_g} \mbox{ (dex/kpc)}.
\end{equation}
Further, we apply the strong-line N2 calibration \citep{pp04} 
recommended by \citet{bresolin12} to our data and plot the 
least-squares fit as a dotted-dashed line in Figure~\ref{fig5}.
Here the N2 calibration is nearly identical 
to the P-method of \citet{pilyugin05}.
The radial oxygen abundance slope of NGC 2403 is robust, 
where the fit to direct and strong-line abundance determinations 
have congruent slopes, but are vertically offset from one another.

\subsubsection{Oxygen Abundance in Context}

The direct abundance measurements of NGC~628 and NGC~2403 
presented here allow comparisons against other spiral galaxies 
with consistent and reliable derivations.
Few previous detailed studies of direct abundances for spiral 
galaxies exist in the literature, and those that do demonstrate a wide range in 
abundance gradient slope: $\sim-0.01-0.04$ dex/kpc.
For example, \citet{bresolin11} found a slope 
of -0.042 dex/kpc for the oxygen abundance gradient in M33, 
in contrast to the measurement of -0.012 dex/kpc by \citet{crockett06}.
Other examples include a slope of -0.023 dex/kpc measured 
for M31 \citep{bresolin12}, -0.011 dex/kpc for NGC 4258 
\citep{bresolin11}, and -0.020 dex/kpc for M51 \citep{bresolin04}.
Thus, in comparison, the gradients measured using the direct method for NGC~628 
(-0.014$\pm$0.002 dex/kpc) and NGC~2403 (-0.028$\pm$0.007 dex/kpc) 
are consistent with studies of other non-barred spiral galaxies (e.g., M33, M31, NGC~4258, and M51).

\begin{figure}[H]
\epsscale{0.6}
\figurenum{5}
\plotone{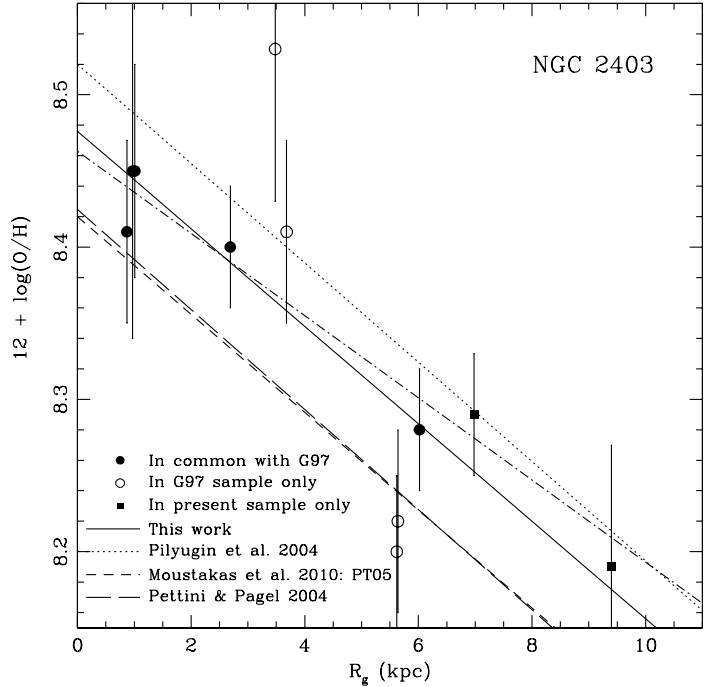}
\caption{Direct oxygen abundances from the MMT
are plotted versus galactocentric radius for NGC 2403. 
The solid circles are observations that overlap with G97 and 
the solid squares are observations of new regions.
Additional direct observations made by G97 are plotted as open circles.
The solid line is a least-squares fit to these data, with little visual or statistical 
dispersion among our data, but some discrepancy with the G97 data.
The dotted dashed line is the fit given for just our data by Equation~\ref{O2403_Us}.
The strong-line oxygen abundance gradient determined by \citet{pilyugin04} 
using the P-method \citep{pilyugin01} is plotted as a dotted line and the gradient determined by
\citet{moustakas10} using the \citet{pilyugin05} P-method as a short dashed line for comparison.
Additionally, we used the N2 calibration of \citet{pp04} to determined the strong-line 
abundance gradient for the present observations and plotted the fit as a long dashed line.}
\label{fig5}
\end{figure}

The new direct radial abundance gradients presented here also highlight  
discrepancies amongst abundance methods when compared to strong-line 
oxygen abundance calibrations.
For NGC~2403, the abundance gradient slope is relatively robust for 
all abundance calibrations shown in Figure~\ref{fig5}. 
However, the strong-line abundance calibrations from the literature
are shifted above and below the direct abundance relationship. 
For NGC~628, we measure a much shallower direct radial abundance gradient 
slope relative to a variety of other methods from the literature.
In particular, previous studies find higher y-intercepts and thus
a significantly higher oxygen abundance predicted for the nucleus of the disk.
Among the observations presented in Figure~\ref{fig4}, the three direct abundances
measured for galactocentric radii less than 5 kpc lie within the vertical scatter 
of the data, but break from the decreasing abundance trend established at larger radii.
These measurements agree with chemical evolution models which predict flatter gradients 
in the dense inner regions of spirals due to the breakdown of the instantaneous 
recycling approximation \citep[e.g.,][]{prantzos00,chiappini03}.
The systematically lower direct oxygen abundances measured for NGC~2403 relative to 
the strong-line P-method gradient of \citet{pilyugin04} and the low direct oxygen abundances
measured for the inner 5 kpcs of NGC~628 could also be due to temperature fluctuations
or gradients.
Such temperature inhomogeneities in metal rich \ion{H}{2} regions have been shown
to cause oxygen abundances to be systematically underestimated by as much
as $\sim0.4$ dex \citep[e.g.,][]{stasinska05,bresolin07}.
As discussed in Section~\ref{sec:temden}, several methods have recently
emerged to correct for these temperature fluctuations \citep[e.g.,][]{pena-guerrero12,nicholls12},
but further work is needed to determine if any are universal solutions to the problem.

%----------------------------------------------------------------------------------------------------------------------
%----------------------------------------------------------------------------------------------------------------------

% Figure 5: NGC 2403 Oxygen Gradient

%----------------------------------------------------------------------------------------------------------------------
%----------------------------------------------------------------------------------------------------------------------

\subsection{Nitrogen}
\label{sec:N}

\citet{garnett90} found a relatively constant relationship in the 
N/O ratio versus O/H for low metallicity star-forming galaxies, 
but with a scatter larger than could be understood in terms 
of observational uncertainties.
He suggested that different delivery times of N and O could explain 
the relatively large scatter seen in N/O at low metallicities.
For example, when a low metallicity galaxy has a burst of 
star formation, contributions from massive stars will drive 
O to higher values, while driving N/O values lower.
Later, contributions from intermediate mass stars 
will raise the N/O while keeping O/H constant.
Then, if many star-forming galaxies at this oxygen 
abundance are measured, but are at various times 
since their burst initiated, a spread in N/O will be seen.
At higher metallicities (i.e., 12+log(O/H) $\ge$ 8.0), secondary 
nitrogen production becomes increasingly significant, 
causing the average N/O to increase with O/H \citep{pagel85}.
We can, therefore, use the radial relationship of N/O in spiral galaxies 
to determine which nucleosynthetic mechanisms are dominant.
We present the absolute and relative nitrogen abundances for 
NGC~628 and NGC~2403 in Tables~\ref{tbl7}-\ref{tbl9}.

\subsubsection{NGC 628}

\begin{figure}[H]
\figurenum{6}
\epsscale{0.6}
\plotone{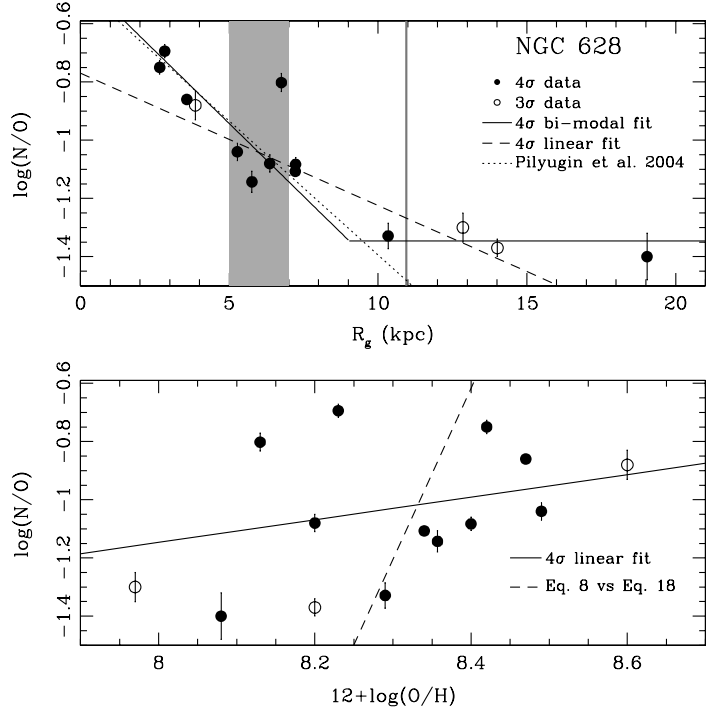}
\caption{Top: The N/O ratio relationship is plotted with galactocentric radius for NGC 628.
The dashed line represents our least-squares fit over the entire 4$\sigma$ dataset.
This linear fit is significantly shallower than the relationship 
determined by \citet{pilyugin04} (dotted line).
However, these data are better fit by a bi-modal relationship (solid lines), where a 
steeper gradient exists in the inner part of the galaxy and flattens out past 10 kpc.
A vertical line has been added to mark the luminosity radius at $R_{25}=10.95$ kpc.
We also highlight the data in the 5 $\lesssim$ $R_{g} \lesssim 7$ range, as it displays
significant scatter relative to the rest of the fit.
Bottom: N/O is plotted over the range in O/H, as is commonly done to look for regions
dominated by different nitrogen production mechanisms.
The solid fit is a simple least-squares linear fit to the data.
The dashed line is derived from the inner galaxy fits in Figure~\ref{fig4} and the upper
panel here. The scatter in this plot suggests large variations in nitrogen
or temperature structure for a given oxygen abundance/region.  }
\label{fig6}
\end{figure}

The upper panel of Figure~\ref{fig6} displays the trend of 
log(N/O) with galactocentric radius in NGC~628.
For N/O, the range of values extend from log(N/O) = $-0.69$ 
at small radii, decreasing outward to $-1.40$ at larger radii.
\citet{pilyugin04} used a single linear fit to characterize the 
N/O relationship with galactocentric radii and found a fairly 
shallow radial gradient (dotted line).
Following this approach, we find a much shallower 
relationship for the 4$\sigma$ data:
\begin{equation}
	\log(\mbox{N/O}) = (-0.49\pm0.08) + (-0.089\pm0.011)\times{R_g}\  \mbox{ (dex/kpc)},
\end{equation}
with a dispersion in log(N/O) of $\sigma$ = 0.11 dex. 
This weighted least squares fit to all 11 4$\sigma$ 
data points is plotted as a dashed line in Figure~\ref{fig6}.

By visual inspection, the data appear to lie in two separate regions.
The range from the inner part of the galaxy out to a galactocentric radius of 
$\approx10$ kpc, near the luminosity radius ($R_{25}$ = 10.95 kpc),
is fit reasonably well by a steeply declining fit (solid line):
\begin{equation}
	\log(\mbox{N/O}) = (-0.45\pm0.08) + (-0.100\pm0.013)\times{R_g}\  \mbox{ (dex/kpc)},
\end{equation}
with a dispersion in log(N/O) of $\sigma$ = 0.12 dex.
This relationship agrees well with the fit found by \citep{pilyugin04}, 
which only sampled the inner galaxy.
Beyond $R_{g}$ = 10 kpc, the outer galaxy appears to flatten out in N/O ratio,
and so can be fit with a constant value of $\log({\mbox{N/O}}) = -1.35$ (solid line),
with a dispersion of $\sigma$ = 0.04 dex.
The \ion{H}{2} regions of NGC 628 measured at distances beyond $R_{25}$,
which have relatively low oxygen abundances (12+log(O/H) $<$ 8.3), 
have an average N/O ratio that is high relative to the plateau seen for 
low metallicity dwarf galaxies (e.g., Garnett 1990: log(N/O) = -1.46; 
van Zee \& Haynes 2006: log(N/O) = -1.41; 
Nava et al.\ 2006: log(N/O) = -1.43; Berg et al.\ 2012: log(N/O) = -1.56).
However, the average value of $\log({\mbox{N/O}}) = -1.35$ measured here  
for $8.0 \lesssim 12+\log({\mbox{O/H}})$ $\lesssim 8.3$ is in agreement 
with the N/O ratios seen for dwarf galaxies of similar oxygen abundances 
in Figure 6 of \cite{berg12}.
An additional point worth noting is the scatter in log(N/O) highlighted in the 
upper panel of Figure~\ref{fig6} for 5 $\lesssim$ $R_{g} \lesssim 7$.
NGC628+185-52 is especially discrepant, deviating by more than 0.2 dex 
from the linear fit.
If this spread is real and not just due to observational uncertainties, 
it could be indicative of local nitrogen pollution by intermediate mass stars.

The lower panel of Figure~\ref{fig6} allows us to further examine this bi-modal 
relationship by fitting the N/O ratio over the range in oxygen abundance.
The clear trend observed versus radius in the upper panel is not 
obvious in the plot versus oxygen abundance.
If we assume the bi-modal relationship is correct, there are three points 
with 12+log(O/H) $<$ 8.25 that have N/O values that are located more than 
5$\sigma$ above the average value of the outer disk in the upper panel.
It is interesting that the three largest outliers from this plot correspond to 
the three measurements of oxygen abundance (at the 4$\sigma$ 
detection level) below the best fit in Figure~4, including NGC+185-52,
which is the highly discrepant point in the upper panel of Figure~\ref{fig6}. 
This could indicate that these regions have oxygen abundances 
that are underestimated, possibly due to temperature inhomogeneities 
within the \ion{H}{2} region \citep{peimbert67} or a non Maxwell-Boltzmann 
electron distribution \citep{nicholls12}.
If we combine the oxygen abundance radial gradient calculated in 
Equation~7 with the inner N/O abundance gradient from the upper panel of Figure~
\ref{fig6} (Equation~16), the result is the steep dashed line in the lower panel.
Based on this relationship, there is a large scatter in 
N/O abundance for a given oxygen abundance.
Instead, the data in the bottom panel of Figure~\ref{fig6} is best fit by a 
standard linear least-squares fit to the entire data range (solid line):
\begin{equation}
	\log(\mbox{N/O}) = -4.25 + 0.388\times{\mbox{12+log(O/H)}},
\end{equation}
with a dispersion in log(N/O) of $\sigma$ = 0.21 dex.
These nitrogen variations offer a different explanation for the divergence of the direct
abundance gradient and the strong-line N2 gradient seen for NGC~628 in Figure~\ref{fig4}. 

Following the method laid out in \citet{zahid11}, we use a likelihood ratio F-test
in order to quantify the significance of the bi-modal model versus the linear model fit.
In the case where ${\chi_1}^2>{\chi_2}^2$, the F-statistic is given by
\begin{equation}
	F = \frac{  \frac{{\chi_1}^2 - {\chi_2}^2} {P_2 - P_1} }{ \frac{{\chi_2}^2} {N - P_2}},
\end{equation}
where ${\chi_1}^2$ corresponds to model 1, which is always the simpler model of 
the two, P is the number of parameters of the fit and N is the number of data points.
Under the null hypothesis that model 2 does not give a significantly better fit than model 
1, F has an F-distribution characterized by $(P_2 - P_1,N - P_2)$ degrees of freedom.
However, if ${\chi_1}^2$ $<$ ${\chi_2}^2$ an alternative F-statistic must be used:
\begin{equation}
	F = \frac{ {{\chi_1}^2}/{dof_1} } { {{\chi_2}^2}/{dof_2} },
\end{equation} 
where the degrees of freedom are given by $dof_i = N - P_i$ such that F
has an F-distribution with $(dof_1,dof_2)$.
In the present case, model 1 is a 2-parameter linear fit, with a ${\chi_1}^2\approx13.3$
and model 2 is a 5-parameter bi-modal fit (each of the two lines has a parameter for
slope and intercept, plus an additional parameter for the location of the break), with a
${\chi_2}^2\approx4.7$. This gives an $F = 3.68$ with (3,6) degrees of freedom.
We then used the IDL routine MPFTEST \citep{markwardt09} to determine that the 
null-hypothesis only has a significance level of 8.2\%. 
This result suggests that model 2 is a significant improvement over model 1 
at the 91.8\% confidence level, where the break occurs near 12+log(O/H) 
= 8.3, which roughly corresponds with $R_g = R_{25}$.  
Thus, we find that while the N/O abundance follows a bi-modal relationship with 
galactocentric distance, significant variations in either temperature or nitrogen abundance 
result in a large scatter for a given oxygen abundance.

%----------------------------------------------------------------------------------------------------------------------

% Figure 6: NGC 628 Nitrogen Gradient

%----------------------------------------------------------------------------------------------------------------------

\subsubsection{NGC 2403}

\begin{figure}[H]
\epsscale{0.6}
\figurenum{7}
\plotone{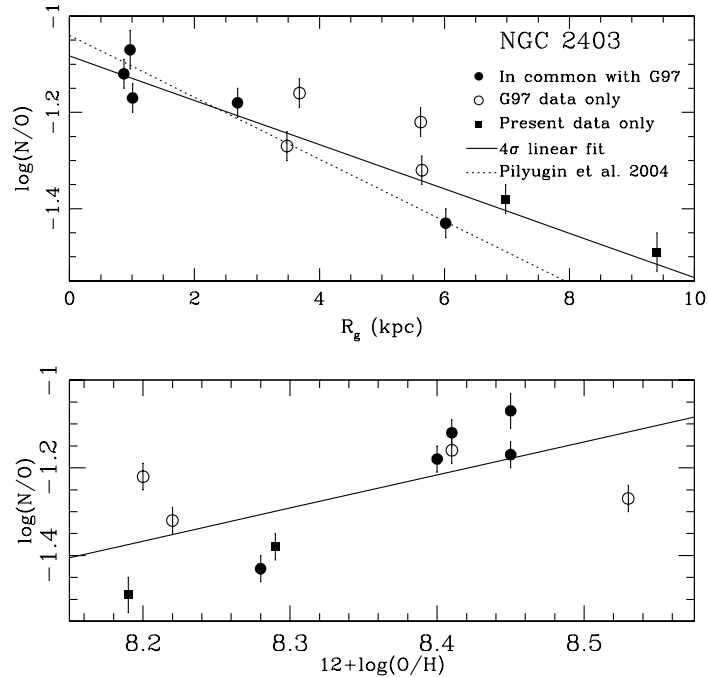}
\caption{
Top: The N/O ratio relationship is plotted with galactocentric radius for NGC 2403.
The solid line represents the least-squares fit to the present data 
plus G97 data over the full radial extent of the observations.
This linear fit demonstrates a somewhat shallower slope than the 
relationship determined by \citet{pilyugin04} (dotted line).
Note that these data only sample the inner disk of NGC~2403 ($R < R_{25}$).
Bottom: The solid line represents the best fit to the data: a simple least-squares linear relationship.}
\label{fig7}
\end{figure}

The top panel of Figure~\ref{fig7} displays the trend of 
log(N/O) with galactocentric radius in NGC~2403.
The gradient in N/O for NGC 2403 is much shallower 
at small radii than in NGC 628.
Near the center, the measured log(N/O) values only 
reach $-1.07$, and then level off at a value of log(N/O) 
= $-1.42$ past $R_g$ = 6 kpc.
If we characterize the combined 11 point data set with a single 
least squares weighted fit (solid line), we find a fit of
\begin{equation}
	\log(\mbox{N/O}) = (-1.08\pm0.02) + (-0.043\pm0.003)\times{R_g}  \mbox{ (dex/kpc)},
\end{equation}
with a dispersion in log(N/O) of $\sigma$ = 0.05 dex. 
\citet{pilyugin04} also used a single linear fit to characterize the N/O relationship 
with galactocentric radii, finding a slightly steeper radial gradient (dotted line):
\begin{equation}
	\log(\mbox{N/O}) = -1.040 - 0.064\times{R_g}  \mbox{ (dex/kpc)}.
\end{equation}

Unlike the trend seen for N/O in NGC~628, 
visually the NGC~2403 data do no follow a bi-modal relationship,
and an F-test confirms this.
However, the luminosity radius for NGC 2403 is $R_{25} = 10.1$ 
kpc, and so the outer optical disk is not sampled by our data set.
If the $R_{25}$ of a galaxy is typically where the break in the N/O ratio 
occurs, then we would not expect to see it in our analysis of NGC 2403, 
and a declining gradient makes sense with respect to our results for NGC 628.
While a general declining trend is observed, note that the vertical spread in the data 
is relatively small compared to the gradient seen in Figure~\ref{fig6} for NGC~628.
At the radial extent of our data, the N/O decline reaches log(N/O)$\sim-1.5$.
Perhaps a plateau in N/O exists near this level farther out in the galaxy (past $R_{25}$), 
similar to the plateau seen for dwarf galaxies \citep[see, e.g.,][]{garnett90,berg12}.

In the lower panel of Figure~\ref{fig7} we have plotted 12+log(O/H) vs.\ 
log(N/O) for a second look at the N/O relationship.
The 4$\sigma$ data are fit best by a simple least-squares fit (solid line):
\begin{equation}
	\log(\mbox{N/O}) =  -7.56 + 0.755\times{\mbox{12+log(O/H)}},
\end{equation}
with a dispersion in log(N/O) of $\sigma$ = 0.10 dex.
Again, it is likely that our data did not sample large enough galactocentric radii
or low enough metallicities to see the plateau related to primary nitrogen production.

\subsubsection{Nitrogen Abundance in Context}

If the trends of increasing N/O ratios towards the center of both galaxies are real, 
it would be indicative of the effects of secondary N production in those regions.
Note that \citet{garnett90} proposed that much of the scatter in the 12+log(O/H) 
vs.\ log(N/O) relationship could be explained by the time delay between producing 
oxygen in massive stars and secondary nitrogen from intermediate mass stars.
Then local nitrogen pollution from intermediate mass stars could provide an 
explanation for the scatter we see in our data in the inner disks.

In the outer disk, many previous studies note no correlation between 
12+log(O/H) and the relative N/O abundance, but do not sample data at radii 
greater than $R_{25}$ \citep[e.g.,][]{bresolin04,bresolin05,bresolin11,zurita12}.
Barred spiral galaxies are known to show a flattening in their gradient correlated with 
the strength of their bar \citep{martin94}.
For example, \citet{zahid11} measured a break in the strongly barred spiral, 
NGC 3359, near 12+log(O/H)$\approx 8.3$,
corresponding to radii between $1\cdot R_{25}$ and $1.5\cdot R_{25}$.
Similarly, \citet{kennicutt03a} found a break near 12+log(O/H)$\approx 8.0$ 
and $R_g = 1\cdot R_{25}$ in M101 and \citet{bresolin09b} showed M83 
exhibits a break near $R_g = 1\cdot R_{25}$ and 12+log(O/H)$\ge 8.3$.
However, a few examples of observed breaks in non-barred spiral 
galaxies also exist in the literature.
 \citet{goddard11} used a variety of strong-line methods in the mixed spiral galaxy, 
 NGC 4625, to show a break near 12+log(O/H) $> 8.3$ at $R_g = R_{25}$ 
 and \citet{bresolin12} measured a break in the non-barred spiral NGC 3621 near 
 $R_g = 1\cdot R_{25}$ for oxygen abundances in the range of $8.3 <$ 12+log(O/H) 
 $< 8.7$ (depending on the abundance method used). 

In the present data of NGC 628, and the examples listed above, a pattern is emerging 
in which N production is transitioning from secondary to primary past $R_{25}$.
Breaks have typically been explained as the result of gas being mixed by radial 
flows associated with a bar \citep[e.g.,][]{vila-costas92,zaritsky94,dutil99}.
However, if breaks are observed to be common to all spiral galaxies, then a 
more general theory will be needed to include non-barred spiral galaxies
(for instance, see \cite{bresolin99} for discussion of other theories).
Furthermore, the break in log(N/O) (near 12+log(O/H) $> 8.3$) is observed at 
a much higher oxygen abundance than has been widely observed in dwarf 
galaxies \citep[see e.g.,][]{garnett90,izotov99,vanzee06a,nava06,berg12}.
If further studies confirm this elevated level for turnover, it would have
the potential to provide new insights into the production of nitrogen
and possible metallicity-dependent yields. 
Note that \citet{pilyugin03} challenged the existence of breaks, showing
that artificial breaks can be observed when the strong-line $R_{23}$ 
method is used to determine abundances.
This result further motivates the need for a more significant sample of 
spiral galaxies with direct abundance gradients and spanning a 
wide range of Hubble types.

For isolated galaxies there is still no strong consensus 
on how metals and gas are transported.
Possible mechanisms of gas mixing include magnetorotational instabilities 
\citep{sellwood99}, gas outflows and infall \citep{santillan07,dalcanton07,zahid12}, 
differential rotation \citep{wada99}, and spiral-bar resonance over-lap \citep{minchev10}.
Cold mode flows have been measured in several spiral galaxies from 
\ion{H}{1} studies \citep[see e.g.,][]{ryan-weber03,elson10}, and are consistent 
with driving central material outward and mixing into the extended disk, 
causing higher-than-expected oxygen abundances in the plateau.
\citet{salim10} and \citet{lemonias11}, among others, have argued that 
cold-mode accretion feeds the ongoing star formation in extended disks, 
but the low metallicity expected for pristine infalling gas contrasts with 
the moderately high oxygen abundances measured in the disk.
Alternatively, \citet{werk11} argue for the hot gas phase mixing 
scenario proposed by \citet{tassis08}, an argument nearly 
identical to those in \citet{ks96} and \citet{ks97}.
However, this model assumes that mergers are the dominant source of 
turbulence, and thus doesn't answer the question of isolated spiral galaxies.
Further, \citet{oppenheimer10} suggest a different process of ``recycled winds", where 
material is driven into the halo by star formation and later reacquired by the galaxy.
This idea is supported by observations from \citet{tumlinson11} of oxygen rich
halos surrounding star-forming galaxies, but which seem to be removed or transformed
during the transition to quiescence. 
As discussed in \citet{bresolin11}, many questions concerning the mechanisms of 
galactic disk assembly and evolution remain and will need to be answered using
ongoing and future work involving optical spectroscopy of faint \ion{H}{2} regions 
out to large galactocentric radii in spiral galaxies.

%----------------------------------------------------------------------------------------------------------------------

% Figure 7: NGC 2403 N Gradient

%----------------------------------------------------------------------------------------------------------------------

\subsection{Sulfur, Neon, and Argon}
\label{sec:alpha}

Stellar nucleosynthesis calculations \citep[e.g.,][]{woosley95} 
indicate that $\alpha$-element and oxygen production occur mainly in 
massive stars of a small mass range, and thus are expected to trace each other closely.
To test this idea in NGC~628 and NGC~2403, we measured the absolute and relative 
sulfur, neon, and argon abundances, which are tabulated in Tables~\ref{tbl7}-\ref{tbl9}.

\subsubsection{NGC 628}
Sulfur, neon, and argon abundances measured for NGC~628 
are plotted versus oxygen abundance in Figure~\ref{fig8}(a).
For each element we have fit the average alpha abundance for the present sample
with a solid line and the standard deviation displayed.
For comparison, we also determined the error weighted least-squares fit using FITEXY,
but found the fits to be no better than the flat relationships.
In addition, we plot strong-line abundances from \citet{vanzee98b} as 
open circles in Figure\ref{fig8}(a).
We fit the combined data set with a simple least-squares linear 
relationship, which was found to be nearly constant in all three cases. 
Even with the addition of the \citet{vanzee98b} data, the 
scatter around a constant value is small compared to 
the large range in oxygen abundance sampled.
Thus, the $\alpha$-elements in NGC 628 investigated here 
can all be described by a constant relationship over oxygen 
abundance such that the $\alpha$-elements behave as 
expected for the nucleosynthetic products of massive stars. 

\subsubsection{NGC 2403}
Sulfur, neon, and argon abundances measured for NGC~2403 
are plotted versus oxygen abundance in Figure~\ref{fig8}(b).
Similar to NGC~628, the alpha elements investigated here are fit well
by an average alpha abundance for the present sample.
The average for each element (solid line) and the standard deviation
from these 0-slope fits are illustrated in Figure~\ref{fig8}(b).
For comparison, we also determined the error weighted least-squares fit using 
FITEXY and found them to be consistent with the flat relationships plotted.
Strong-line abundances are plotted from G97 as crosses 
and from \citep{vanzee98b} as open circles, showing that the data from 
the present sample lies with in the scatter seen around a constant value.
These additional data, especially in the case of argon, help to sample a more uniform 
distribution over the range of oxygen abundance. 
We fit the combined data set with a simple least-squares linear 
relationship, which was found to be nearly constant in all three cases. 
The fact that we are not able to distinguish a trend as more significant 
than the constant average value indicates that these elements are 
likely a consequence of massive star nucleosynthesis.

\subsubsection{$\alpha$-Elements in Context}

In a comprehensive study of M33, \citet{kwitter81} found that Ne, 
N, S, and Ar gradients followed that which they derived for oxygen. 
From stellar nucleosynthesis calculations \citep[e.g.,][]{woosley95}, 
oxygen and neon both seem to be produced mainly in stars larger 
than 10 solar masses, and thus are expected to trace each other closely.
Similarly, \citet{henry90} finds that Ne/O is constant over 
a wide range of O/H for planetary nebulae.
Thus, the present gradient analysis for NGC~628 and NGC~2403 
agree with previous findings that $\alpha$/O remains
constant over the range in oxygen abundance.

In contrast to this viewpoint, \citet{willner02} derived a neon gradient that 
is significantly shallower than the oxygen gradient observed in M33. 
While sulfur is also traditionally assumed to have a constant S/O ratio 
\citep{garnett89}, some specific cases, such as the work of \citet{vilchez88} 
on M33, find a slower decline of sulfur than oxygen with radius.
Thus, a larger sample of galaxies are needed to determine whether constant
$\alpha$-element abundance gradients are a universal trend amongst spiral galaxies.

\begin{figure}[H]
\figurenum{8}
\begin{center}
\begin{tabular}{cc}
	\resizebox{85mm}{!}{\includegraphics{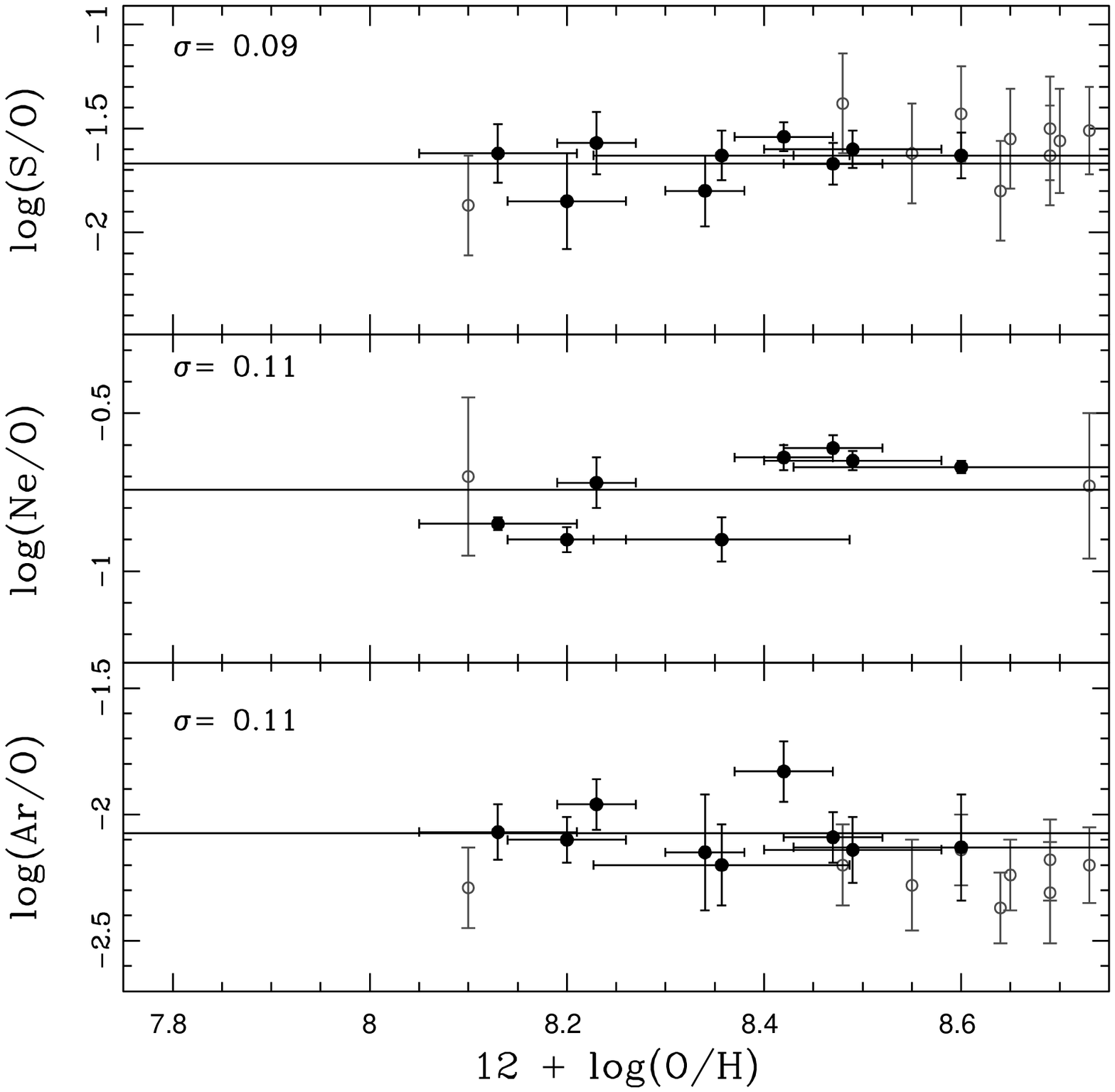}} & \resizebox{85mm}{!}{\includegraphics{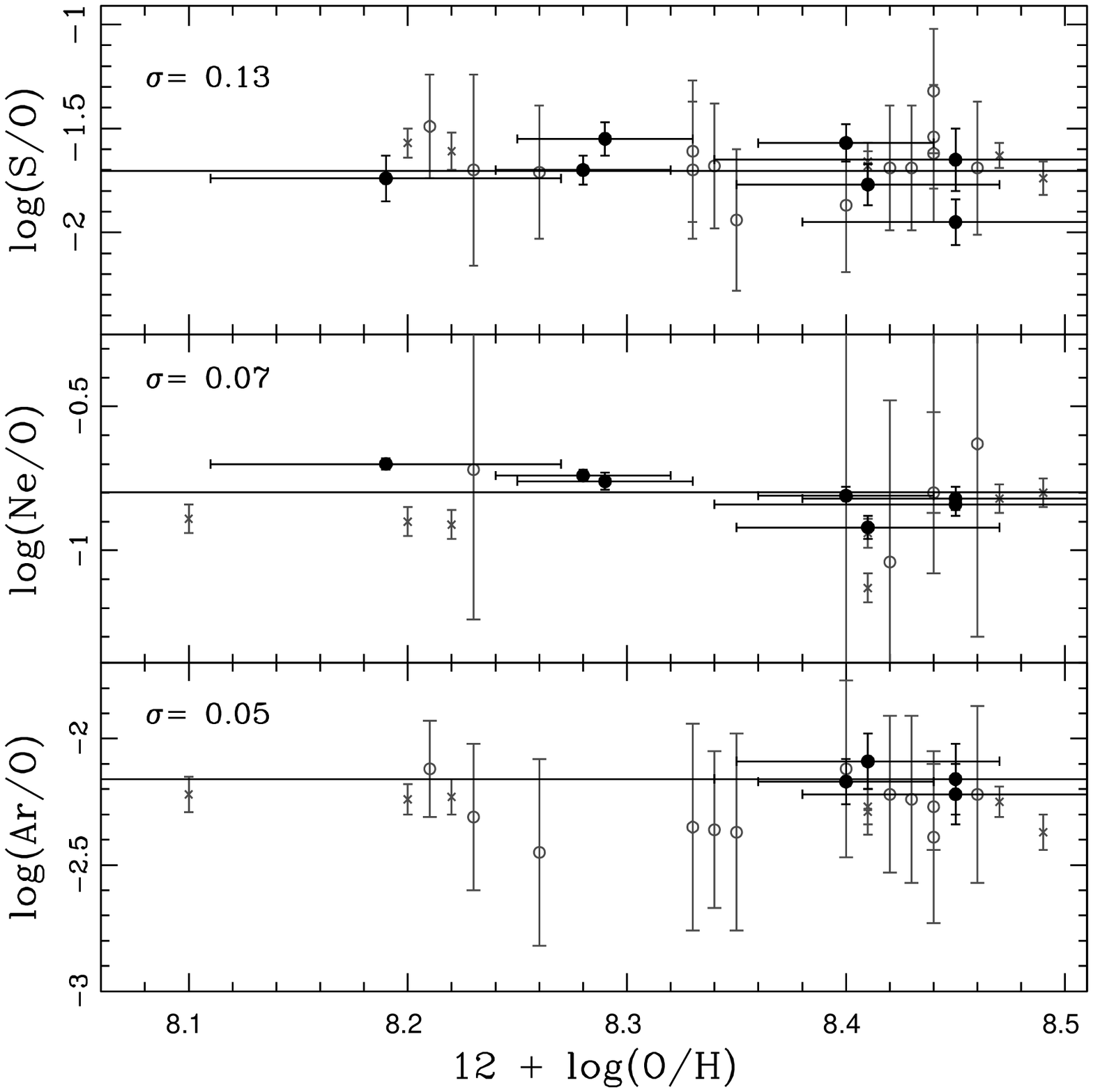}} \\
\end{tabular}
\end{center}
\caption{On the left, $\alpha$-elements are plotted versus oxygen abundance 
for NGC 628 and on the right, $\alpha$-elements are plotted for NGC 2403.
The present data are plotted as solid points and fit with $\alpha$-element averages (solid lines).
 Data from G97 (crosses) and from \citet[][open circles]{vanzee98b} are plotted
 in comparison, helping to extend the coverage over oxygen abundance.
 This larger data set generally follows the constant $\alpha$/O abundance vs. O
 for each element.}
\label{fig8}
\end{figure}

The conclusions drawn from Tables~\ref{tbl8}-\ref{tbl9} and Figure~\ref{fig8} 
support the convention that $\alpha$-element abundances (S, Ar, and Ne)
and O evolve in lockstep for both NGC~628 and NGC~2403. 
The final adopted fits to the abundance gradients for NGC~628 and
NGC~2403 are given in Table~\ref{tbl100}.

%----------------------------------------------------------------------------------------------------------------------

% Figure 8: NGC 628 and NGC 2403 Alpha Gradients

%----------------------------------------------------------------------------------------------------------------------
%----------------------------------------------------------------------------------------------------------------------

% Table 100: Galaxy Fits
\begin{deluxetable}{ccclcc}
\tabletypesize{\footnotesize}
\tablewidth{0pt}
\tablecaption{Adopted Abundance Gradients for NGC~628 and NGC~2403}
\tablehead{
\multicolumn{1}{c}{y} & \CH{} 	& \CH{x} 	& \CH{Equation of Correlation}  & \CH{$\sigma$}	& \CH{Portion of Disk}	}
\startdata
\multicolumn{6}{c}{NGC~628} \\
\hline \\			
{12+log(O/H)} 	& {vs.}		& {$R_g$}			& {y = (8.43$\pm$0.03) + (-0.017$\pm0.002)\times$x}		& {0.10} & {Total}	\\
{log(N/O)}		& {vs.}		& {$R_g$}			& {y = (-0.45$\pm$0.08) + (-0.100$\pm0.013)\times$x}		& {0.11} & {Inner} 	\\
{}			& {}			& {}				& {y = -1.35}										& {0.04} & {Outer} 	\\
{log(N/O)} 	& {vs.} 		& {12+log(O/H)} 	& {y = -4.25 + 0.388$\times$x}							& {0.21} & {Total} 	\\
{log(S/O)} 		& {vs.} 		& {12+log(O/H)}	& {y = -1.67}										& {0.09} & {Total} 	\\
{log(Ne/O)} 	& {vs.} 		&{12+log(O/H)}		& {y = -0.74}										& {0.11} & {Total} 	\\
{log(Ar/O)} 	& {vs.} 		& {12+log(O/H)}	& {y = -2.07}										& {0.11} & {Total} 	\\
\hline \\
\multicolumn{6}{c}{NGC~2403} \\
\hline \\			
{12+log(O/H)} 	& {vs.}		& {$R_g$}			& {y = (8.48$\pm$0.04) + (-0.032$\pm0.007)\times$x}		& {0.07} & {Total}	\\
{log(N/O)}		& {vs.}		& {$R_g$}			& {y = (-1.08$\pm$0.02) + (-0.043$\pm0.004)\times$x}		& {0.05} & {Total} 	\\
{log(N/O)} 	& {vs.} 		& {12+log(O/H)} 	& {y = -7.56 + 0.755$\times$x}							& {0.10} & {Total} 	\\
{log(S/O)} 		& {vs.} 		& {12+log(O/H)}	& {y = -1.67}										& {0.09} & {Total} 	\\
{log(Ne/O)} 	& {vs.} 		&{12+log(O/H)}		& {y = -0.74}										& {0.11} & {Total} 	\\
{log(Ar/O)} 	& {vs.} 		& {12+log(O/H)}	& {y = -2.07}										& {0.11} & {Total} 	\\
\enddata
\tablecomments{The adopted best fits to the abundance gradients measured for the 4$\sigma$ NGC~628 data 
and the combined sample of present observations plus G97 for NGC~2403.}
\label{tbl100}
\end{deluxetable}

%----------------------------------------------------------------------------------------------------------------------
%----------------------------------------------------------------------------------------------------------------------

\section{CONCLUSIONS}
\label{sec:conclusion}

Spiral galaxies pose a challenge to abundance work, as a single 
abundance measurement is not sufficient to characterize the entire galaxy. 
Thus, high quality spectra of many \ion{H}{2} regions that enable direct 
abundances are required to securely measure the oxygen abundance gradient.

We have uniformly determined oxygen abundance metallicities for 14 \ion{H}{2} 
regions in the local spiral galaxy NGC 628 and seven \ion{H}{2} regions in NGC 2403. 
With high-quality spectroscopic observations, we measured the intrinsically 
faint [\ion{O}{3}] $\lambda4363$ and/or [\ion{N}{2}] $\lambda5755$ auroral 
lines at  a strength of 3$\sigma$ or greater, and explicitly determine 
electron temperatures in all 14 \ion{H}{2} regions in NGC 628.
The seven \ion{H}{2} regions in NGC 2403 were chosen to have strong
auroral emission lines and, therefore, all have electron temperatures
determined at a significance above 4$\sigma$.

Our results provide the first ever oxygen abundance gradient 
derived solely from direct abundance measurements
for \ion{H}{2} regions in NGC 628.
From the $4\sigma$ data we derive an oxygen abundance 
gradient of 12 + log(O/H) = (8.43$\pm$0.03) + 
(-0.017$\pm$0.002)$\times{R_g}$ (dex/kpc),
with a dispersion in log(O/H) of $\sigma$ = 0.10 dex. 
This is a significantly shallower slope than found 
by previous empirical abundance studies.
As all previous studies of NGC 2403 have relied primarily 
on strong-line abundances, with only a small number of 
direct abundances, the seven direct oxygen abundances 
presented here allow an improved metallicity gradient analysis.
The result from these H II regions, which extend from a galactocentric 
radius of 0.96 to 10.4 kpc, is an oxygen abundance gradient of 
12 + log(O/H) = $(8.48\pm0.04) + (-0.032\pm0.007)\times{R_g}$ 
(dex/kpc), with a modest dispersion of $\sigma = 0.07$ dex.

Additionally, we measure the absolute and 
relative N, S, Ne, and Ar Abundances.
For the N/O ratio we find a negative gradient with 
increasing galactocentric radius in the inner disks 
of both NGC 628 and NGC 2403. 
Further out in the outer disk, this relationship flattens 
out in NGC 628, where we find a plateau past $R_g = 
R_{25}$ near an oxygen abundance of 12+log(O/H) = 8.3. 
Similar trends are seen in other spiral galaxies, where the 
$R_{25}$ radius and oxygen abundances of 12+log(O/H) 
= 8.3 and greater mark points of transition in nitrogen production.
NGC 2403, and other previously studied 
galaxies, are best fit with a single linear fit.
However, in theses cases, the galaxies do not have observations 
in their extended disks beyond $R_{25}$ and therefore their 
mechanisms of nitrogen production are not readily 
comparable to galaxies with observations of greater radial coverage.
 
As expected for $\alpha$-process elements, S/O, Ne/O, and 
Ar/O appear to be constant over a range in oxygen abundance 
in  both NGC 628 and NGC 2403 such that the 
$\alpha$-elements and O are produced in lock-step.
Since there is a general paucity of abundance data from 
individual spiral galaxies, more of these accurate \ion{H}{2} 
region datasets are necessary to understand individual 
galaxy processes and relative trends among galaxies.

%----------------------------------------------------------------------------------------------------------------------
%----------------------------------------------------------------------------------------------------------------------

\acknowledgements
We are grateful to the referee for a quick and insightful report which
substantially improved the organization and scientific impact of the paper.
We are also grateful to Liese van Zee for helpful discussion and the use
of her H$\alpha$ and R-band imaging. 
DAB is grateful for support from a Penrose Fellowship, a NASA Space Grant 
Fellowship, and a Dissertation Fellowship from the University of Minnesota.
EDS is grateful for partial support from the University of Minnesota and NSF Grant AST-1109066.
Special thanks to Henry Lee for his role helping to prepare the Gemini observations.

Observations reported here were obtained at the MMT Observatory, a 
joint facility of the Smithsonian Institution and the University of Arizona.
MMT observations were obtained as part of the University of Minnesota's 
guaranteed time on Steward Observatory facilities through membership in 
the Research Corporation and its support for the Large Binocular Telescope.
This work is based, in part, on observations obtained at the Gemini Observatory, 
which is operated by the Association of Universities for Research in Astronomy, 
Inc., under a cooperative agreement with the NSF on behalf of the Gemini 
partnership: the National Science Foundation (United States), the National 
Research Council (Canada), CONICYT (Chile), the Australian Research 
Council (Australia), Ministério da Ciência, Tecnologia e Inovação (Brazil) 
and Ministerio de Ciencia, Tecnología e Innovación Productiva (Argentina).
The authors wish to recognize and acknowledge the very significant cultural 
role and reverence that the summit of Mauna Kea has always had within 
the indigenous Hawaiian community.  
We are most fortunate to have the opportunity to conduct observations from this mountain.

This research has made use of NASA's Astrophysics Data System
Bibliographic Services and the NASA/IPAC Extragalactic Database
(NED), which is operated by the Jet Propulsion Laboratory, California
Institute of Technology, under contract with the National Aeronautics
and Space Administration.
This work was initiated as part of the Spitzer Space Telescope Legacy Science
Program and was supported by National Aeronautics and Space Administration 
(NASA) through contract 1336000 issued by the Jet Propulsion Laboratory (JPL), 
California Institute of Technology (Caltech) under NASA contract 1407.

%----------------------------------------------------------------------------------------------------------------------
%----------------------------------------------------------------------------------------------------------------------

\bibliography{mybib}{}

\newpage 

%----------------------------------------------------------------------------------------------------------------------

%----------------------------------------------------------------------------------------------------------------------

\textheight=10.5in

% Table 4:
\begin{deluxetable}{lccccc}
\tabletypesize{\scriptsize}
\tablenum{4.5}
\setlength{\tabcolsep}{0.04in} 
\tablecaption{ Emission-Line Intensities and Equivalent Widths for Gemini Observations of \ion{H}{2} regions in NGC 628}
\tablewidth{0pt}
\tablehead{
\CH{} & \multicolumn{5}{c}{$I(\lambda)/I(\mbox{H}\beta)$} }
\startdata	
%{Ion}					& {338}			& {65}			& {299}			& {311}			& {180}			\\
{Ion}						& {+9+76}			& {-76-29}			& {-59+84}		& {+82-74}		& {-134+69}		\\
\hline
{[O II] $\lambda$3727}		& 1.78$\pm$0.07	& 1.66$\pm$0.06	& 2.34$\pm$0.05	& 1.83$\pm$0.07	& 2.48$\pm$0.09	\\
%{H10 $\lambda$3798}		& \nodata			& \nodata			& \nodata			& \nodata			& \nodata			\\
{H9 $\lambda$3835}		& 0.061$\pm$0.005	& 0.03$\pm$0.01	& \nodata			& \nodata			& \nodata			\\
%{[Ne III] $\lambda$3868}	& \nodata			& \nodata			& \nodata			& \nodata			& \nodata			\\
{He I+H8 $\lambda$3889} 	& 0.23$\pm$0.01	& 0.22$\pm$0.01	& 0.32$\pm$0.01	& 0.12$\pm$0.01	& 0.17$\pm$0.01	\\
{[Ne III]+H7 $\lambda$3968}	& 0.16$\pm$0.01	& 0.13$\pm$0.01	& 0.15$\pm$0.01	& 0.11$\pm$0.01	& 0.14$\pm$0.01	\\
%{He I $\lambda$4026}		& \nodata			& \nodata			& \nodata			& \nodata			& \nodata			\\
%{[S II] $\lambda$4068}		& \nodata			& \nodata			& \nodata			& \nodata			& \nodata			\\
{H$\delta$ $\lambda$4101}	& 0.28$\pm$0.01	& 0.27$\pm$0.01	& 0.26$\pm$0.01	& 0.27$\pm$0.01	& 0.27$\pm$0.01	\\
{H$\gamma$ $\lambda$4340}	& 0.50$\pm$0.02	& 0.53$\pm$0.01	& 0.47$\pm$0.01	& 0.49$\pm$0.01	& 0.46$\pm$0.01	\\
{[O III] $\lambda$4363}		& \nodata			& \nodata			& \nodata			& \nodata			& \nodata			\\
{He I $\lambda$4471}		& 0.023$\pm$0.003	& 0.023$\pm$0.002	& 0.035$\pm$0.03	& \nodata			& 0.022$\pm$0.002	\\
%{[Fe III] $\lambda$4658}		& \nodata			& \nodata			& \nodata			& \nodata			& \nodata			\\
%{He II $\lambda$4686}		& \nodata			& \nodata			& \nodata			& \nodata			& \nodata			\\
{H$\beta$ $\lambda$4861}	& 1.00$\pm$0.02	& 1.00$\pm$0.02	& 1.00$\pm$0.02	& 1.00$\pm$0.02	& 1.00$\pm$0.02	\\
%{He I $\lambda$4921}		& \nodata			& \nodata			& \nodata			& \nodata			& \nodata			\\
{[O III] $\lambda$4959}		& 0.069$\pm$0.001	& 0.086$\pm$0.002	& 0.23$\pm$0.01	& 0.083$\pm$0.002	& 0.231$\pm$0.005	\\
{[O III] $\lambda$5007}		& 0.183$\pm$0.004	& 0.26$\pm$0.01	& 0.67$\pm$0.01	& 0.239$\pm$0.004	& 0.68$\pm$0.01	\\
{N I $\lambda$5199}			& 0.013$\pm$0.001	& 0.020$\pm$0.001	& 0.013$\pm$0.001	& \nodata			& 8.2$\pm$0.8E-4	\\
%{[Fe III] $\lambda$5271}		& \nodata			& \nodata			& \nodata			& \nodata			& \nodata			\\
%{[Cl III] $\lambda$5518}		& \nodata			& \nodata			& \nodata			& \nodata			& \nodata			\\
%{[Cl III] $\lambda$5538}		& \nodata			& \nodata			& \nodata			& \nodata			& \nodata			\\
{[N II] $\lambda$5755}		& 6.7$\pm$0.5E-3	& 7.7$\pm$0.5E-3	& 0.007$\pm$0.001	& 4.8$\pm$1.6E-3	& 0.005$\pm$0.001	\\
{He I $\lambda$5876}		& 0.086$\pm$0.001	& 0.083$\pm$0.002	& 0.136$\pm$0.003	& 0.084$\pm$0.002	& 0.09$\pm$0.002	\\
{[O I] $\lambda$6300}		& 0.017$\pm$0.001	& 0.027$\pm$0.001	& 0.024$\pm$0.001	& 0.018$\pm$0.001	& 0.016$\pm$0.001	\\
{[S III] $\lambda$6312}		& 3.5$\pm$0.4E-3	& 0.003$\pm$0.001	& 0.006$\pm$0.001	& 0.001$\pm$0.001	& 0.008$\pm$0.001	\\
{[O I] $\lambda$6363}		& 6.0$\pm$0.4E-3	& 0.007$\pm$0.001	& 0.007$\pm$0.001	& 0.007$\pm$0.001	& 0.006$\pm$0.001	\\
{[N II] $\lambda$6548}		& 0.34$\pm$0.01	& 0.30$\pm$0.01	& 0.31$\pm$0.01	& 0.30$\pm$0.01	& 0.21$\pm$0.01	\\
{H$\alpha$ $\lambda$6563}	& 2.98$\pm$0.10	& 2.99$\pm$0.10	& 2.97$\pm$0.06	& 2.94$\pm$0.11	& 2.85$\pm$0.10	\\
{[N II] $\lambda$6584}		& 1.02$\pm$0.04	& 0.93$\pm$0.03	& 0.92$\pm$0.02	& 0.88$\pm$0.03	& 0.64$\pm$0.02	\\
{He I $\lambda$6678}		& 0.026$\pm$0.001	& 0.024$\pm$0.001	& 0.030$\pm$0.001	& 0.025$\pm$0.001	& 0.027$\pm$0.001	\\
{[S II] $\lambda$6717}		& 0.31$\pm$0.01	& 0.37$\pm$0.01	& 0.31$\pm$0.01	& 0.33$\pm$0.01	& 0.25$\pm$0.01	\\
{[S II] $\lambda$6731}		& 0.22$\pm$0.01	& 0.27$\pm$0.01	& 0.24$\pm$0.01	& 0.25$\pm$0.01	& 0.18$\pm$0.01	\\
{He I $\lambda$7065}		& 0.012$\pm$0.001	& 0.011$\pm$0.001	& 0.020$\pm$0.001	& 0.012$\pm$0.001	& 0.015$\pm$0.001	\\
{[Ar III] $\lambda$7136}		& 0.040$\pm$0.002	& 0.033$\pm$0.001	& 0.066$\pm$0.001	& 0.033$\pm$0.001	& 0.060$\pm$0.003	\\
{[C II] $\lambda$7236}		& 4.5$\pm$0.3E-3	& \nodata			& \nodata			& \nodata			& \nodata			\\
{He I $\lambda$7281}		& 4.4$\pm$0.3E-3	& \nodata			& \nodata			& 0.005$\pm$0.001	& 3.3$\pm$0.4E-3	\\
{[O II] $\lambda$7320}		& 10.0$\pm$0.5E-3	& \nodata			& 0.018$\pm$0.001	& 0.013$\pm$0.001	& 0.018$\pm$0.001	\\
{[O II] $\lambda$7330}		& 6.9$\pm$0.4E-3	& \nodata			& 0.016$\pm$0.001	& 0.009$\pm$0.001	& 0.011$\pm$0.001	\\						
\hline	
					
{C(H$\beta$)}				& 0.71$\pm$0.04	& 0.55$\pm$0.04   	& 0.72$\pm$0.01	& 0.58$\pm$0.04   	& 0.30$\pm$0.04   	\\
{$F$(H$\beta$)}			& 73.9$\pm$0.5   	& 88.9$\pm$0.8   	& 52.1$\pm$0.0	& 57.7$\pm$0.2    	& 75.9$\pm$0.5    	\\
{EW(H$\beta$)}			& 226 			& 90.9          		& 143			& 44.5			& 67.1          		\\
{EW(H$\alpha$)}			& 1340          		& 622			& 420			& 361			& 550.          		\\
																		
\tablebreak %------------------------------------------------------------------------------------------------------------------------------------------------------------------------------------------------------------------------------------------------------------------------------------

% {ion}					& {447}			& {485}     			& {220}			& {128}			& {434}			\\
{Ion}						& {-44-159}		& {-2+182}  		& {+185-52}		& {-190+80}		& {-90+186}		\\
\hline
{[O II] $\lambda$3727}	       	& \nodata			& 2.32$\pm$0.09 	& 1.67$\pm$0.06	& 2.73$\pm$0.10	& 2.82$\pm$0.06	\\
{H10 $\lambda$3798}	      	& \nodata			& \nodata   		& \nodata			& \nodata			& 0.029$\pm$0.003	\\
{H9 $\lambda$3835}	      	& \nodata			& 0.08$\pm$0.01 	& 0.11$\pm$0.004	& 0.062$\pm$0.003	& \nodata			\\
{[Ne III] $\lambda$3868}      	& 0.16$\pm$0.01	& 0.06$\pm$0.01 	& \nodata			& 0.050$\pm$0.003	& 0.073$\pm$0.003	\\
{He I+H8 $\lambda$3889}       	& 0.19$\pm$0.01	& 0.21$\pm$0.01 	& 0.31$\pm$0.01	& 0.21$\pm$0.01	& 0.188$\pm$0.003	\\
{[Ne III]+H7 $\lambda$3968}   	& 0.15$\pm$0.01	& 0.022$\pm$0.005 	& 0.19$\pm$0.01	& 0.19$\pm$0.01	& 0.173$\pm$0.001	\\
{[S II] $\lambda$4068}	      	& \nodata			& \nodata			& \nodata     		& 0.017$\pm$0.002	& 0.009$\pm$0.001	\\
{H$\delta$ $\lambda$4101}     	& 0.26$\pm$0.01	& 0.26$\pm$0.01 	& 0.27$\pm$0.01	& 0.267$\pm$0.007	& 0.26$\pm$0.01	\\
{H$\gamma$ $\lambda$4340}	& 0.51$\pm$0.01	& 0.51$\pm$0.01 	& 0.49$\pm$0.01	& 0.48$\pm$0.01	& 0.48$\pm$0.01	\\
{[O III] $\lambda$4363}	      	& 0.015$\pm$0.003	& \nodata   		& \nodata			& 7.4$\pm$0.8E-3	& 0.010$\pm$0.001 \\
{He I $\lambda$4471}	      	& 0.034$\pm$0.002	& 0.032$\pm$0.002 	& 0.026$\pm$0.004	& 0.028$\pm$0.002	& 0.035$\pm$0.001	\\
{[Fe III] $\lambda$4658}      	& \nodata			& \nodata     		& \nodata			& \nodata			& 0.011$\pm$0.001	\\
{He II $\lambda$4686}	      	& \nodata			& \nodata     		& \nodata			& \nodata			& 0.049$\pm$0.001	\\
{H$\beta$ $\lambda$4861}      	& 1.00$\pm$0.02	& 1.00$\pm$0.02	& 1.00$\pm$0.02	& 1.00$\pm$0.02	& 1.00$\pm$0.02	\\
{He I $\lambda$4921}	      	& \nodata			& 0.009$\pm$0.001	& \nodata			& \nodata			& 8.8$\pm$0.0E-3	\\
{[O III] $\lambda$4959}	      	& 0.72$\pm$0.01	& 0.46$\pm$0.01 	& 0.30$\pm$0.01	& 0.39$\pm$0.01	& 0.53$\pm$0.01	\\
{[O III] $\lambda$5007}	      	& 2.17$\pm$0.04	& 1.33$\pm$0.03 	& 0.88$\pm$0.02	& 1.20$\pm$0.02	& 1.56$\pm$0.03	\\
{N I $\lambda$5199}	      		& \nodata			& 0.007$\pm$0.001 	& 0.012$\pm$0.001	& 0.01$\pm$0.01	& 8.4$\pm$0.3E-3	\\
{[Fe III] $\lambda$5271}      	& \nodata			& \nodata     		& \nodata			& \nodata			& 2.2$\pm$0.4E-3	\\
{[Cl III] $\lambda$5518}      	& 0.005$\pm$0.001	& 3.9$\pm$0.6E-3 	& \nodata			& \nodata			& \nodata			\\
{[Cl III] $\lambda$5538}      	& \nodata			& 2.4$\pm$0.6E-3 	& \nodata			& 2.7$\pm$0.4E-3	& \nodata			\\
{[N II] $\lambda$5755}	      	& 0.004$\pm$0.001	& 0.005$\pm$0.001 	& 0.007$\pm$0.001	& 5.3$\pm$0.3E-3	& 5.5$\pm$0.4E-3	\\
{He I $\lambda$5876}	      	& 0.126$\pm$0.003	& 0.111$\pm$0.002 	& 0.106$\pm$0.003	& 0.103$\pm$0.003	& 0.131$\pm$0.003	\\
{[O I] $\lambda$6300}	      	& 0.021$\pm$0.001	& 0.016$\pm$0.001	& 0.022$\pm$0.001	& 0.032$\pm$0.001	& 19.5$\pm$0.4E-3	\\
{[S III] $\lambda$6312}	      	& 0.009$\pm$0.001	& 0.009$\pm$0.001	& \nodata			& 9.0$\pm$0.3E-3	& 11.5$\pm$0.4E-3	\\
{[O I] $\lambda$6363}	      	& 0.007$\pm$0.001	& 0.005$\pm$0.001 	& 0.006$\pm$0.001	& 0.011$\pm$0.001	& 6.2$\pm$0.4E-3	\\
{[N II] $\lambda$6548}	      	& 0.14$\pm$0.01	& 0.139$\pm$0.005 	& 0.20$\pm$0.01	& 0.19$\pm$0.01	& 0.175$\pm$0.004	\\
{H$\alpha$ $\lambda$6563}     & 2.97$\pm$0.10	& 2.91$\pm$0.10 	& 2.89$\pm$0.10	& 2.93$\pm$0.10	& 2.89$\pm$0.06	\\
{[N II] $\lambda$6584}	      	& 0.39$\pm$0.01	& 0.41$\pm$0.01 	& 0.60$\pm$0.02	& 0.59$\pm$0.02	& 0.52$\pm$0.01	\\
{He I $\lambda$6678}	      	& 0.038$\pm$0.001	& 0.033$\pm$0.001 	& 0.032$\pm$0.001	& 0.031$\pm$0.001	& 0.031$\pm$0.001	\\
{[S II] $\lambda$6717}	      	& 0.22$\pm$0.01	& 0.19$\pm$0.01 	& 0.29$\pm$0.011	& 0.34$\pm$0.01	& 0.235$\pm$0.005	\\
{[S II] $\lambda$6731}	      	& 0.16$\pm$0.01	& 0.135$\pm$0.005 	& 0.20$\pm$0.007	& 0.25$\pm$0.01	& 0.171$\pm$0.003	\\
{He I $\lambda$7065}	      	& 0.021$\pm$0.001	& 0.017$\pm$0.001 	& 0.014$\pm$0.001	& \nodata			& 19.7$\pm$0.4E-3	\\
{[Ar III] $\lambda$7136}        	& 0.092$\pm$0.004	& 0.075$\pm$0.003 	& 0.060$\pm$0.003	& \nodata			& 0.081$\pm$0.002	\\
%{[C II] $\lambda$7236}	      	& \nodata			& \nodata			& \nodata			& \nodata			& \nodata			\\
{He I $\lambda$7281}	      	& 0.005$\pm$0.001	& 4.6$\pm$0.4E-3 	& \nodata			& \nodata			& \nodata			\\
{[O II] $\lambda$7320}	      	& 0.017$\pm$0.001	& 0.018$\pm$0.001 	& \nodata			& \nodata			& \nodata			\\
{[O II] $\lambda$7330}	      	& 0.014$\pm$0.001	& 0.014$\pm$0.001 	& \nodata			& \nodata			& \nodata			\\
\hline															
{C(H$\beta$)}				& 0.36$\pm$0.04	& 0.22$\pm$0.04 	& 0.33$\pm$0.04   	& 0.32$\pm$0.04 	& 0.34$\pm$0.01	\\
{$F$(H$\beta$)}			& 212$\pm$0		& 57.9$\pm$0.2 	& 47.3$\pm$0.9    	& 75.7$\pm$0.5   	& 220.$\pm$0		\\
{EW(H$\beta$)}			& 71				& 111   			& 155            		& 174			& 153			\\
{EW(H$\alpha$)}			& 516			& 897   			& 1137          		& 1350 			& 1120			\\
\enddata
\tablecomments{Optical line fluxes for \ion{H}{2} regions measured from Gemini 
NGC 628 spectra using deblended Gaussian fits and multiple component fits when necessary. 
Fluxes are relative to H$\beta$ = 1.00 and are corrected for reddening. 
The H$\beta$ flux is given for reference, with units of $10^{-16}$ erg s$^{-1}$ cm$^{-2}$. 
EWs are given in units of \AA.
Note that uncertainties listed in this table reflect the statistical uncertainties in the flux 
through the slit only, and do not account for slit losses.} 
\label{tbl4}
\end{deluxetable}

% Table 5:
\begin{deluxetable}{lccccc}
\tabletypesize{\scriptsize}
\tablenum{4.6}
\setlength{\tabcolsep}{0.04in} 
\tablecaption{ Emission-Line Intensities and Equivalent Widths for MMT Observations of \ion{H}{2} regions in NGC 628}
\tablewidth{0pt}
\tablehead{
\CH{} & \multicolumn{4}{c}{$I(\lambda)/I(\mbox{H}\beta)$} }
\startdata	
%{ion}					& {vZ6}  			& {C-4}			& {E-6}			& {F-1}			\\
{ion}						& {+295-16} 		& {-277+240}		& {+186+355}		& {+503+208}		\\	
\hline																			
{[O II] $\lambda$3727}		& 1.59$\pm$0.12	& 3.10$\pm$0.06	& 2.61$\pm$0.08	& 2.81$\pm$0.10	\\
{H12 $\lambda$3750}		& 0.026$\pm$0.004	& \nodata			& 0.036$\pm$0.001	& 0.026$\pm$0.008 \\
{H11 $\lambda$3771}		& 0.035$\pm$0.004	& \nodata			& 0.041$\pm$0.001	& 0.035$\pm$0.008	\\
{H10 $\lambda$3798}		& 0.046$\pm$0.005	& \nodata			& 0.05$\pm$0.01	& 0.05$\pm$0.01	\\
{He I $\lambda$3820}		& 0.009$\pm$0.004	& \nodata			& \nodata			& \nodata			\\
{H9 $\lambda$3835}		& 0.068$\pm$0.006	& \nodata			& 0.09$\pm$0.01	& 0.06$\pm$0.01	\\
{[Ne III] $\lambda$3868}		& 0.35$\pm$0.02	& 0.14$\pm$0.02	& 0.31$\pm$0.01	& 0.27$\pm$0.01	\\
{He I+H8 $\lambda$3889} 	& 0.19$\pm$0.01	& 0.24$\pm$0.02	& 0.20$\pm$0.01	& 0.19$\pm$0.01	\\
{[Ne III]+H7 $\lambda$3968}	& 0.26$\pm$0.02	& 0.25$\pm$0.02	& 0.26$\pm$0.01	& 0.25$\pm$0.01	\\
{He I $\lambda$4026}		& 0.016$\pm$0.002	& \nodata			& \nodata			& 0.022$\pm$0.006	\\
{[S II] $\lambda$4068}		& 0.009$\pm$0.002	& \nodata			& \nodata     		& 0.013$\pm$0.006 \\
{H$\delta$ $\lambda$4101}	& 0.25$\pm$0.01	& 0.26$\pm$0.02	& 0.28$\pm$0.01	& 0.27$\pm$0.01	\\
{H$\gamma$ $\lambda$4340}	& 0.47$\pm$0.02	& 0.47$\pm$0.02	& 0.47$\pm$0.01	& 0.45$\pm$0.01	\\
{[O III] $\lambda$4363}		& 0.031$\pm$0.002	& 0.035$\pm$0.011	& 0.029$\pm$0.008	& 0.036$\pm$0.002	\\
{He I $\lambda$4471}		& 0.038$\pm$0.002	& \nodata			& 0.041$\pm$0.007	& 0.040$\pm$0.005	\\
{H$\beta$ $\lambda$4861}	& 1.00$\pm$0.02	& 1.00$\pm$0.02	& 1.00$\pm$0.01	& 1.00$\pm$0.02	\\
{He I $\lambda$4921}		& 0.010$\pm$0.001	& \nodata			& \nodata			& 0.010$\pm$0.003 \\
{[O III] $\lambda$4959}		& 1.52$\pm$0.03	& 0.65$\pm$0.01	& 1.21$\pm$0.02	& 1.14$\pm$0.02	\\
{[O III] $\lambda$5007}		& 4.53$\pm$0.10	& 1.94$\pm$0.03	& 3.64$\pm$0.07	& 3.43$\pm$0.07	\\
{He I $\lambda$5015}		& 0.018$\pm$0.001	& \nodata			& \nodata			& 0.019$\pm$0.003 \\
{[Cl III] $\lambda$5518}		& 0.004$\pm$0.001	& \nodata			& \nodata			& \nodata			\\
{He I $\lambda$5876}		& 0.12$\pm$0.01	& 0.05$\pm$0.02	& 0.095$\pm$0.005	& 0.10$\pm$0.01	\\
{[O I] $\lambda$6300}		& 0.016$\pm$0.001	& \nodata			& 0.019$\pm$0.004	& 0.027$\pm$0.005	\\
{[S III] $\lambda$6312}		& 0.015$\pm$0.001	& \nodata			& 0.016$\pm$0.004	& 0.012$\pm$0.005	\\
{[O I] $\lambda$6363}		& 0.005$\pm$0.001	& \nodata			& 0.010$\pm$0.004	& 0.010$\pm$0.005	\\
{[N II] $\lambda$6548}		& 0.038$\pm$0.003	& 0.07$\pm$0.01	& 0.058$\pm$0.005	& 0.054$\pm$0.006	\\
{H$\alpha$ $\lambda$6563}	& 2.91$\pm$0.10	& 2.86$\pm$0.06	& 2.91$\pm$0.08	& 2.84$\pm$0.010	\\
{[N II] $\lambda$6584}		& 0.12$\pm$0.01	& 0.24$\pm$0.01	& 0.175$\pm$0.006	& 0.16$\pm$0.03	\\
{He I $\lambda$6678}		& 0.034$\pm$0.009	& 0.029$\pm$0.010	& 0.029$\pm$0.004	& 0.032$\pm$0.004	\\
{[S II] $\lambda$6717}		& 0.099$\pm$0.008	& 0.25$\pm$0.01	& 0.16$\pm$0.02	& 0.18$\pm$0.01	\\
{[S II] $\lambda$6731}		& 0.070$\pm$0.005	& 0.16$\pm$0.01	& 0.10$\pm$0.01	& 0.12$\pm$0.01	\\
\hline																																								
{C(H$\beta$)}				& 0.27$\pm$0.10	& 0.06$\pm$0.01      	& 0.39$\pm$0.03	& 0.18$\pm$0.04	\\
{$F$(H$\beta$)}			& 33.4$\pm$0.7	& 4.62$\pm$0.08   	& 7.29$\pm$0.04	& 10.7$\pm$0.2	\\
{EW(H$\beta$)}			& 228			& 68.8   			& 161			& 369			\\
{EW(H$\alpha$)}			& 1220			& 412            		& 944			& 2020			\\
\enddata
\tablecomments{Optical line fluxes for \ion{H}{2} regions measured from MMT 
NGC 628 spectra using deblended Gaussian fits and multiple component fits when necessary. 
Fluxes are relative to H$\beta$ = 1.00 and are corrected for reddening. 
The H$\beta$ flux is given for reference, with units of $10^{-16}$ erg s$^{-1}$ cm$^{-2}$. 
EWs are given in units of \AA.
Note that uncertainties listed in this table reflect the statistical uncertainties in the flux 
through the slit only, and do not account for slit losses.} 
\label{tbl5}
\end{deluxetable}
																	%------------------------------------------------------------------------------------------------------------------------------------------------------------------------------------------------------------------------------------------------------------------------------------

% Table 6:
\begin{deluxetable}{lccccccc}
\tabletypesize{\scriptsize}
\tablenum{4.7}
\setlength{\tabcolsep}{0.04in} 
%\tablecaption{ Emission-Line Intensities and Equivalent Widths for MMT Observations of \ion{H}{2} regions in NGC 2403}
\tablecaption{ Emission-Line Intensities for MMT Observations of NGC 2403}
\tablewidth{0pt}
\tablehead{
\CH{} & \multicolumn{6}{c}{$I(\lambda)/I(\mbox{H}\beta)$} }
\startdata										
%{ion}					& {35}			& {24}			& {38}			& {44}			& {9}				& {376}			& {423}			\\ 
{ion}						& {-7+36}			& {-30+45}		& {+13+31}		& {+104+24}		& {-133-146}		& {+376-106}		& {-423-10}		\\
\hline	
{[O II] $\lambda$3727}		& 2.82$\pm$0.12	& 1.90$\pm$0.08	& 2.34$\pm$0.10	& 2.47$\pm$0.10	& 2.44$\pm$0.11	& 2.44$\pm$0.11	& 2.31$\pm$0.10	\\
{H12 $\lambda$3750}		& 0.018$\pm$0.002	& \nodata			& 0.020$\pm$0.002	& 0.027$\pm$0.001	& 0.020$\pm$0.001	& \nodata			& 0.021$\pm$0.001	\\
{H11 $\lambda$3771}		& 0.030$\pm$0.002	& \nodata			& 0.025$\pm$0.002	& 0.032$\pm$0.002	& 0.030$\pm$0.002	& 0.050$\pm$0.005	& 0.030$\pm$0.002	\\
{H10 $\lambda$3798}		& 0.047$\pm$0.002	& \nodata			& 0.036$\pm$0.002	& 0.042$\pm$0.002	& 0.041$\pm$0.002	& 0.050$\pm$0.004	& 0.040$\pm$0.002	\\
{He I $\lambda$3820}		& \nodata			& \nodata			& \nodata			& \nodata			& 0.008$\pm$0.001	& \nodata			& \nodata			\\
{H9 $\lambda$3835}		& 0.075$\pm$0.004	& \nodata			& 0.061$\pm$0.003	& 0.065$\pm$0.002	& 0.062$\pm$0.003	& 0.063$\pm$0.004	& 0.065$\pm$0.003	\\
{[Ne III] $\lambda$3868}		& 0.046$\pm$0.002	& 0.038$\pm$0.002	& 0.062$\pm$0.003	& 0.090$\pm$0.004	& 0.25$\pm$0.01	& 0.18$\pm$0.01	& 0.29$\pm$0.01	\\
{He I+H8 $\lambda$3889} 	& 0.20$\pm$0.01	& 0.26$\pm$0.01	& 0.18$\pm$0.01	& 0.18$\pm$0.01	& 0.18$\pm$0.01	& 0.19$\pm$0.01	& 0.18$\pm$0.01	\\
{[Ne III]+H7 $\lambda$3968}	& 0.18$\pm$0.01	& 0.17$\pm$0.01	& 0.14$\pm$0.01	& 0.18$\pm$0.01	& 0.21$\pm$0.01	& 0.21$\pm$0.01	& 0.21$\pm$0.01	\\
{He I $\lambda$4026}		& 0.015$\pm$0.001	& \nodata			& 0.008$\pm$0.002	& 0.013$\pm$0.001	& 0.015$\pm$0.001	& \nodata			& 0.012$\pm$0.001 	\\
{[S II] $\lambda$4068}		& 0.021$\pm$0.002	& \nodata			& 0.012$\pm$0.002	& 0.014$\pm$0.002	& 0.019$\pm$0.001	& \nodata			& 0.020$\pm$0.002	\\
{H$\delta$ $\lambda$4101}	& 0.29$\pm$0.01	& 0.26$\pm$0.01	& 0.26$\pm$0.01	& 0.26$\pm$0.01	& 0.26$\pm$0.01	& 0.26$\pm$0.01	& 0.27$\pm$0.01	\\
{H$\gamma$ $\lambda$4340}	& 0.52$\pm$0.02	& 0.47$\pm$0.01	& 0.48$\pm$0.01	& 0.48$\pm$0.01	& 0.47$\pm$0.01	& 0.48$\pm$0.01	& 0.48$\pm$0.01	\\
{[O III] $\lambda$4363}		& 5.5$\pm$1.2E-3	& \nodata			& 0.007$\pm$0.002	& 7.7$\pm$0.5E-3	& 0.034$\pm$0.001	& 0.019$\pm$0.002	& 0.038$\pm$0.001 	\\
{He I $\lambda$4387}		& \nodata			& \nodata			& \nodata			& \nodata			& 3.5$\pm$0.9E-3	& \nodata			& 0.003$\pm$0.001	\\
{He I $\lambda$4471}		& 0.040$\pm$0.001	& 0.018$\pm$0.002	& 0.031$\pm$0.002	& 2.7$\pm$0.5E-3	& 0.040$\pm$0.001	& 0.035$\pm$0.003	& 0.035$\pm$0.001	\\	
{[Fe III] $\lambda$4658}		& \nodata			& \nodata			& \nodata			& 8.0$\pm$0.4E-3	& 6.0$\pm$0.7E-3	& \nodata			& 5.8$\pm$0.7E-3	\\
{[Ar IV] + HeI $\lambda$4713}	& \nodata			& \nodata			& \nodata			& \nodata			& 5.0$\pm$0.7E-3	& \nodata			& 4.5$\pm$0.7E-3	\\
{H$\beta$ $\lambda$4861}	& 1.00$\pm$0.02	& 1.00$\pm$0.02	& 1.00$\pm$0.02	& 1.00$\pm$0.02	& 1.00$\pm$0.02	& 1.00$\pm$0.02	& 1.00$\pm$0.02	\\
{He I $\lambda$4921}		& \nodata			& \nodata			& 6.8$\pm$1.0E-3	& 10.9$\pm$0.4E-3	& 0.010$\pm$0.001	& 0.010$\pm$0.002	& 8.5$\pm$0.8E-3	\\
{[O III] $\lambda$4959}		& 0.43$\pm$0.01	& 0.333$\pm$0.008	& 0.43$\pm$0.01	& 0.65$\pm$0.01	& 1.27$\pm$0.03	& 1.09$\pm$0.03	& 1.33$\pm$0.03	\\
{[O III] $\lambda$5007}		& 1.26$\pm$0.03	& 1.00$\pm$0.01	& 1.28$\pm$0.03	& 1.96$\pm$0.05	& 3.78$\pm$0.09	& 3.25$\pm$0.07	& 4.00$\pm$0.09	\\
{He I $\lambda$5015}		& 0.023$\pm$0.001	& 0.014$\pm$0.002	& 0.020$\pm$0.001	& 0.024$\pm$0.001	& 0.023$\pm$0.001	& \nodata			& 0.028$\pm$0.001	\\
{N I $\lambda$5199}			& 0.009$\pm$0.001	& 0.006$\pm$0.002	& 6.2$\pm$1.0E-3	& 9.2$\pm$0.4E-3	& 0.004$\pm$0.001	& 0.005$\pm$0.002	& 5.9$\pm$0.7E-3	\\
{[Fe III] $\lambda$5271}		& \nodata			& \nodata			& \nodata			& 3.8$\pm$0.3E-3	& 2.8$\pm$0.7E-3	& \nodata			& \nodata			\\
{[Cl III] $\lambda$5518}		& 3.5$\pm$0.7E-3	& 5.3$\pm$0.9E-3	& 3.4$\pm$1.0E-3	& 3.9$\pm$0.3E-3	& 4.1$\pm$0.8E-3	& \nodata			& 3.7$\pm$0.7E-3	\\
{[Cl III] $\lambda$5538}		& 2.2$\pm$0.7E-3	& \nodata			& 2.4$\pm$1.0E-3	& 2.7$\pm$0.3E-3	& 2.7$\pm$0.7E-3	& \nodata			& 2.9$\pm$0.7E-3	\\
{[N II] $\lambda$5755}		& 5.2$\pm$0.6E-3	& 3.7$\pm$0.9E-3	& 3.5$\pm$0.4E-3	& 4.2$\pm$0.5E-3	& 2.4$\pm$0.3E-3	& \nodata			& 2.2$\pm$0.7E-3	\\
{He I $\lambda$5876}		& 0.117$\pm$0.004	& 0.107$\pm$0.004	& 0.106$\pm$0.003	& 0.116$\pm$0.004	& 0.116$\pm$0.004	& 0.129$\pm$0.005	& 0.109$\pm$0.004	\\
{[O I] $\lambda$6300}		& 0.013$\pm$0.001	& 8.8$\pm$0.9E-3	& 9.0$\pm$0.5E-3	& 0.020$\pm$0.001	& 0.028$\pm$0.001	& 0.006$\pm$0.002	& 0.023$\pm$0.001	\\
{[S III] $\lambda$6312}		& 0.008$\pm$0.001	& 6.0$\pm$0.9E-3	& 6.0$\pm$0.4E-3	& 11.6$\pm$0.4E-3	& 0.015$\pm$0.001	& 0.014$\pm$0.002	& 0.013$\pm$0.001	\\
{[O I] $\lambda$6363}		& 4.2$\pm$0.5E-3	& 2.56$\pm$0.9E-3	& 3.5$\pm$0.4E-3	& 6.6$\pm$0.3E-3	& 9.3$\pm$0.5E-3	& 0.005$\pm$0.002	& 7.4$\pm$0.8E-3	\\
{[N II] $\lambda$6548}		& 0.18$\pm$0.01	& 0.15$\pm$0.01	& 0.15$\pm$0.01	& 0.124$\pm$0.005	& 0.039$\pm$0.002	& 0.060$\pm$0.003	& 0.044$\pm$0.002	\\
{H$\alpha$ $\lambda$6563}	& 2.87$\pm$0.12	& 2.93$\pm$0.12	& 2.88$\pm$0.12	& 2.91$\pm$0.12	& 2.83$\pm$0.12	& 2.88$\pm$0.12	& 2.83$\pm$0.12	\\
{[N II] $\lambda$6584}		& 0.52$\pm$0.02	& 0.45$\pm$0.02	& 0.46$\pm$0.02	& 0.37$\pm$0.02	& 0.18$\pm$0.01	& 0.17$\pm$0.01	& 0.13$\pm$0.01	\\
{He I $\lambda$6678}		& 0.032$\pm$0.001	& 0.027$\pm$0.001	& 0.026$\pm$0.001	& 0.032$\pm$0.001	& 0.029$\pm$0.001	& 0.030$\pm$0.002	& 0.025$\pm$0.001	\\
{[S II] $\lambda$6717}		& 0.24$\pm$0.01	& 0.22$\pm$0.01	& 0.23$\pm$0.01	& 0.21$\pm$0.01	& 0.18$\pm$0.01	& 0.15$\pm$0.01	& 0.131$\pm$0.006	\\
{[S II] $\lambda$6731}		& 0.17$\pm$0.01	& 0.16$\pm$0.01	& 0.16$\pm$0.01	& 0.16$\pm$0.01	& 0.14$\pm$0.01	& 0.111$\pm$0.005	& 0.096$\pm$0.004	\\
{He I $\lambda$7065}		& 0.020$\pm$0.004	& 0.017$\pm$0.003	& 0.014$\pm$0.001	& 0.026$\pm$0.001	& \nodata	 		& \nodata			& \nodata			\\
{[Ar III] $\lambda$7136}		& 0.089$\pm$0.008	& 0.073$\pm$0.005	& 0.061$\pm$0.004	& 0.093$\pm$0.005	& \nodata	 		& \nodata	 		& \nodata			\\
{[O II] $\lambda$7325}		& 0.029$\pm$0.005	& 0.027$\pm$0.005	& 0.020$\pm$0.004	& 0.048$\pm$0.003	& \nodata	 		& \nodata			& \nodata			\\
\hline																																				
{C(H$\beta$)}				& 0.58$\pm$0.05 	& 0.10$\pm$0.05   	& 0.02$\pm$0.05   	& 0.17$\pm$0.05	& 0.06$\pm$0.05	& 0.36$\pm$0.05   	& 0.15$\pm$0.05	\\
{EW(H$\beta$)}			& 248			& 25.4            		& 60.9            		& 164			& 165 			& 103			& 172			\\
{EW(H$\alpha$)}			& 1394 			& 186            		& 284			& 1050			& 953            		& 941			& 883			\\
\enddata
\\[-0.5cm]
\tablecomments{Optical line fluxes for \ion{H}{2} regions measured from the MMT NGC 
2403 spectra using deblended Gaussian fits and multiple component fits when necessary. 
Fluxes are relative to H$\beta$ = 1.00 and are corrected for reddening. 
The H$\beta$ flux is given for reference, with units of $10^{-16}$ erg s$^{-1}$ cm$^{-2}$. 
EWs are given in units of \AA.
Note that uncertainties listed in this table reflect the statistical uncertainties in the flux 
through the slit only, and do not account for slit losses.} 
\label{tbl6}
\end{deluxetable}

%----------------------------------------------------------------------------------------------------------------------

% Table 7: 
\begin{deluxetable}{lccccc}
\tablenum{4.8}
\tabletypesize{\scriptsize}
\setlength{\tabcolsep}{0.05in} 
\tablecaption{Ionic and Total Abundances for Gemini Observations of NGC 628 \label{tbl7}}
\tablewidth{0pt}	
\tablehead{\vspace{-0.6cm}}
\startdata		
%{H$\alpha$ Region}			& {338}			& {65}			& {299}			& {311} 03/16/13	& {180}			\\
{H$\alpha$ Region} 				& {+9+76}			& {-76-29}			& {-59+84}		& {+82-74}		& {-134+69}		\\
\hline					
t(O$_2$)$_{measured}$ (K)		& 8400$\pm$200	& \nodata			& 9500$\pm$100	& 9000$\pm$200	& 8800$\pm$200	\\
t(O$_3$)$_{measured}$ (K)		& \nodata			& \nodata			& \nodata			& \nodata			& \nodata			\\
t(N$_2$)$_{measured}$ (K)		& 7900$\pm$200	& 8500$\pm$200	& 8300$\pm$200	& 7400$\pm$800	& 8300$\pm$500	\\
t(S$_3$)$_{measured}$ (K)		& \nodata			& \nodata			& \nodata			& \nodata			& \nodata			\\		
\newline \\					
t(O$_2$)$_{used}$ (K)        		& 7900$\pm$200	& 8500$\pm$200	& 8300$\pm$200	& 7400$\pm$800	& 8300$\pm$200	\\
t(O$_3$)$_{used}$ (K)			& 7000$\pm$500	& 7800$\pm$500	& 7600$\pm$500	& 6300$\pm$700	& 7600$\pm$400	\\
t(N$_2$)$_{used}$ (K)			& 7900$\pm$200	& 8500$\pm$200	& 8300$\pm$200	& 7400$\pm$800	& 8300$\pm$500	\\
t(S$_3$)$_{used}$ (K)			& 7500$\pm$500	& 8200$\pm$500	& 8000$\pm$500	& 6900$\pm$700	& 8000$\pm$500	\\
\\					
O$^+$/H$^+$ ($\times10^5$)  		& 23.0$\pm$3.1	& 14.4$\pm$1.7	& 22.3$\pm$3.2	& 33.0$\pm$19.4	& 23.3$\pm$7.2	\\
O$^{++}$/H$^{+}$ ($\times10^5$)  	& 3.19$\pm$1.04	& 2.60$\pm$0.68	& 7.59$\pm$2.09	& 6.51$\pm$0.23	& 7.61$\pm$0.22	\\
O/H ($\times10^5$)				& 26.2$\pm$3.3	& 17.0$\pm$1.8	& 29.8$\pm$3.8	& 39.5$\pm$19.4	& 30.9$\pm$7.2	\\
12 + log(O/H)$_{D}$ (dex)		& 8.42$\pm$0.05	& 8.23$\pm$0.04	& 8.47$\pm$0.05	& 8.60$\pm$0.17\D	& 8.49$\pm$0.09	\\
12 + log(O/H)$_{P}$ (dex)		& 8.37$\pm$0.05	& 8.12$\pm$0.04	& 8.06$\pm$0.03	& 8.27$\pm$0.04	& 8.11$\pm$0.04	\\
\\					
N$^{+}$/H$^{+}$ ($\times10^6$)	& 40.6$\pm$3.7	& 29.0$\pm$2.4	& 30.5$\pm$0.28	& 42.8$\pm$15.0	& 21.0$\pm$4.0	\\
N ICF						& 1.139$\pm$0.172	& 1.180$\pm$0.143	& 1.341$\pm$0.168	& 1.197$\pm$0.695 	& 1.326$\pm$0.331 	\\
log(N/O) (dex)					& -0.75$\pm$0.02	& -0.69$\pm$0.02	& -0.86$\pm$0.02	& -0.88$\pm$0.05	& -1.04$\pm$0.03	\\
N/H ($\times10^6$)				& 46.6$\pm$6.4	& 34.5$\pm$4.1	& 41.3$\pm$0.6	& 51.5$\pm$25.9	& 27.9$\pm$6.8	\\
12 + log(N/H) (dex)				& 7.67$\pm$0.06	& 7.54$\pm$0.05	& 7.62$\pm$0.05	& 7.71$\pm$0.18	& 7.45$\pm$0.10	\\
\\					
S$^{+}$/H$^{+}$ ($\times10^7$)	& 24.8$\pm$2.3	& 23.8$\pm$2.0	& 21.5$\pm$0.2	& 32.9$\pm$11.3	& 16.9$\pm$3.2	\\
S$^{++}$/H$^{+}$ ($\times10^7$)	& 36.5$\pm$15.0	& 18.3$\pm$7.7	& 44.9$\pm$16.1	& 22.6$\pm$20.9	& 53.4$\pm$18.5	\\
S ICF						& 0.995$\pm$0.044	& 1.005$\pm$0.052	& 1.058$\pm$0.088	& 1.010$\pm$0.276	& 1.054$\pm$0.176	\\
S/O							& 0.023$\pm$0.007	& 0.025$\pm$0.006	& 0.024$\pm$0.007	& 0.014$\pm$0.010	& 0.024$\pm$0.009	\\
log(S/O) (dex)					& -1.63$\pm$0.11	& -1.60$\pm$0.09	& -1.63$\pm$0.12	& -1.85$\pm$0.23	& -1.62$\pm$0.14	\\
S/H ($\times10^6$)				& 6.10$\pm$1.55	& 4.23$\pm$0.83	& 7.02$\pm$1.73	& 5.60$\pm$2.83 	& 7.41$\pm$2.25	\\
12 + log(S/H) (dex)				& 6.78$\pm$0.10	& 6.63$\pm$0.08	& 6.85$\pm$0.10	& 6.75$\pm$0.18	& 6.87$\pm$0.12	\\
\\					
Ne$^{++}$/H$^{+}$ ($\times10^6$)	& \nodata			& \nodata			& \nodata			& \nodata			& \nodata			\\
Ne ICF						& \nodata			& \nodata			& \nodata			& \nodata			& \nodata			\\
Ne/O							& \nodata			& \nodata			& \nodata			& \nodata			& \nodata			\\
log(Ne/O) (dex)					& \nodata			& \nodata			& \nodata			& \nodata			& \nodata			\\
Ne/H ($\times10^5$)				& \nodata			& \nodata			& \nodata			& \nodata			& \nodata			\\
12 + log(Ne/H) (dex)				& \nodata			& \nodata			& \nodata			& \nodata			& \nodata			\\
\\					
Ar$^{++}$/H$^{+}$ ($\times10^7$)	& 8.34$\pm$1.87	& 5.21$\pm$0.97	& 11.3$\pm$2.2	& 8.87$\pm$2.13	& 10.2$\pm$1.9	\\
Ar ICF						& 4.683$\pm$0.368	& 3.578$\pm$0.258	& 2.155$\pm$0.116	& 3.282$\pm$0.228	& 2.216$\pm$0.122 	\\
log(Ar/O) (dex)					& -1.83$\pm$0.12	& -1.96$\pm$0.10	& -2.09$\pm$0.10	& -2.13$\pm$0.21	& -2.14$\pm$0.13	\\
Ar/H ($\times10^6$)				& 3.90$\pm$1.05	& 1.87$\pm$0.42	& 2.43$\pm$0.56	& 2.91$\pm$0.84	& 2.25$\pm$0.50	\\
12 + log(Ar/H) (dex)				& 6.59$\pm$0.10	& 6.27$\pm$0.09	& 6.39$\pm$0.09	& 6.46$\pm$0.11	& 6.35$\pm$0.09	\\

\tablebreak %------------------------------------------------------------------------------------------------------------------------------------------------------------------------------------------------------------------------------------------------------------------------------------

%{H$\alpha$ Region}			& {447}			& {485}			& {220} 			& {128}			& {434}			\\
{H$\alpha$ Region} 				& {-44-159}		& {-2+182}  		& {+185-52}		& {-190+80}		& {-90+186}		\\
\hline						
t(O$_2$)$_{measured}$ (K)		& \nodata			& 9300$\pm$200	& \nodata			& \nodata			& \nodata			\\
t(O$_3$)$_{measured}$ (K)		& 10100$\pm$800	& \nodata			& \nodata			& 9900$\pm$800	& 10000$\pm$400	\\
t(N$_2$)$_{measured}$ (K)		& 8800$\pm$900	& 9700$\pm$500	& 9300$\pm$700	& 8700$\pm$200	& 9200$\pm$200	\\
t(S$_3$)$_{measured}$ (K)		& \nodata			& \nodata			& \nodata			& \nodata			& \nodata			\\
\newline \\						
t(O$_2$)$_{used}$ (K)        		& 8800$\pm$900	& 9700$\pm$500	& 9300$\pm$700	& 8700$\pm$200	& 9200$\pm$200	\\
t(O$_3$)$_{used}$ (K)			& 10100$\pm$800	& 9600$\pm$500	& 9000$\pm$400	& 9900$\pm$800	& 10000$\pm$400	\\
t(N$_2$)$_{used}$ (K)			& 8800$\pm$900	& 9700$\pm$500	& 9300$\pm$700	& 8700$\pm$200	& 9200$\pm$200	\\
t(S$_3$)$_{used}$ (K)			& 10100$\pm$800	& 9700$\pm$500	& 9200$\pm$500	& 9900$\pm$800	& 10000$\pm$400	\\
\\						
O$^+$/H$^+$ ($\times10^5$)  		& 15.0$\pm$7.6	& 10.0$\pm$2.2	& 8.80$\pm$2.88	& 20.7$\pm$2.4	& 16.3$\pm$2.0	\\
O$^{++}$/H$^{+}$ ($\times10^5$)  	& 7.72$\pm$1.94	& 5.74$\pm$0.16	& 4.76$\pm$1.36	& 4.56$\pm$1.21	& 5.81$\pm$0.80	\\
O/H ($\times10^5$)				& 22.7$\pm$7.8	& 15.8$\pm$2.2	& 13.6$\pm$2.9	& 25.2$\pm$2.7	& 22.1$\pm$2.2	\\
12 + log(O/H)$_{D}$ (dex)		& 8.36$\pm$0.13	& 8.20$\pm$0.06	& 8.13$\pm$0.08	& 8.40$\pm$0.04	& 8.34$\pm$0.04	\\
12 + log(O/H)$_{P}$ (dex)		& 7.02$\pm$0.03	& 7.87$\pm$0.03	& 7.67$\pm$0.03	& 8.03$\pm$0.03	& 8.00$\pm$0.02	\\
\\						
N$^{+}$/H$^{+}$ ($\times10^6$)	& 10.9$\pm$3.3	& 8.39$\pm$1.19	& 13.7$\pm$2.8	& 16.9$\pm$0.1	& 12.7$\pm$0.99	\\
N ICF						& 1.514$\pm$0.479	& 1.572$\pm$0.201	& 1.541$\pm$0.301	& 1.221$\pm$0.145	& 1.357$\pm$0.128	\\
log(N/O) (dex)					& -1.14$\pm$0.04	& -1.08$\pm$0.03	& -0.80$\pm$0.03	& -1.08$\pm$0.02	& -1.11$\pm$0.02	\\
N/H ($\times10^6$)				& 16.4$\pm$5.8	& 13.2$\pm$0.2	& 2.14$\pm$0.48	& 20.8$\pm$2.5	& 17.2$\pm$1.7	\\
12 + log(N/H) (dex)				& 7.21$\pm$0.13	& 7.12$\pm$0.06	& 7.33$\pm$0.09	& 7.32$\pm$0.05	& 7.24$\pm$0.04	\\
\\						
S$^{+}$/H$^{+}$ ($\times10^7$)	& 12.7$\pm$3.8	& 8.23$\pm$1.15	& 13.5$\pm$2.7	& 20.4$\pm$1.7	& 11.9$\pm$0.9	\\
S$^{++}$/H$^{+}$ ($\times10^7$)	& 20.1$\pm$6.8	& 25.3$\pm$0.6	& \nodata			& 21.7$\pm$7.5	& 26.9$\pm$5.8	\\
S ICF						& 1.112$\pm$0.252	& 1.127$\pm$0.103	& \nodata			& 1.018$\pm$0.063	& 1.064$\pm$0.070	\\
S/O							& 0.016$\pm$0.007	& 0.024$\pm$0.006	& \nodata			& 0.017$\pm$0.004	& 0.019$\pm$0.003	\\
log(S/O) (dex)					& -1.80$\pm$0.17	& -1.62$\pm$0.09	& \nodata			& -1.77$\pm$0.09	& -1.73$\pm$0.07	\\
S/H ($\times10^6$)				& 3.64$\pm$1.13	& 3.78$\pm$0.70	& \nodata			& 4.28$\pm$0.81	& 4.13$\pm$0.64	\\
12 + log(S/H) (dex)				& 6.56$\pm$0.12	& 6.58$\pm$0.07	& \nodata			& 6.63$\pm$0.08	& 6.62$\pm$0.06	\\
\\						
Ne$^{++}$/H$^{+}$ ($\times10^6$)	& 17.4$\pm$5.5	& 7.31$\pm$1.20	& \nodata			& 5.80$\pm$1.97	& 8.24$\pm$0.35	\\
Ne ICF						& 2.947$\pm$0.479	& 2.750$\pm$0.201	& \nodata			& 5.533$\pm$0.145	& 3.798$\pm$0.128	\\
Ne/O							& 0.23$\pm$0.02	& 0.13$\pm$0.02	& \nodata			& 0.13$\pm$0.01	& 0.14$\pm$0.01	\\
log(Ne/O) (dex)					& -0.65$\pm$0.03	& -0.90$\pm$0.07	& \nodata			& -0.90$\pm$0.04	& -0.85$\pm$0.02	\\
Ne/H ($\times10^5$)				& 5.12$\pm$1.81	& 2.01$\pm$0.44	& \nodata			& 3.21$\pm$0.47	& 3.13$\pm$0.32	\\
12 + log(Ne/H) (dex)				& 7.71$\pm$0.13	& 7.30$\pm$0.09	& \nodata			& 7.51$\pm$0.06	& 7.50$\pm$0.04	\\
\\						
Ar$^{++}$/H$^{+}$ ($\times10^7$)	& 8.33$\pm$1.58	& 7.55$\pm$1.01	& 6.88$\pm$1.00	& \nodata			& 7.53$\pm$0.92	\\
Ar ICF						& 1.734$\pm$0.073	& 1.661$\pm$0.066	& 1.697$\pm$0.757	& \nodata			& 2.093$\pm$0.109	\\
log(Ar/O) (dex)					& -2.20$\pm$0.16	& -2.10$\pm$0.09	& -2.07$\pm$0.11	& \nodata			& -2.15$\pm$0.07	\\
Ar/H ($\times10^6$)				& 1.44$\pm$0.33	& 1.25$\pm$0.20	& 1.17$\pm$0.20	& \nodata			& 1.58$\pm$0.23	\\
12 + log(Ar/H) (dex)				& 6.16$\pm$0.09	& 6.10$\pm$0.06	& 6.07$\pm$0.07	& \nodata			& 6.20$\pm$0.06	\\
\enddata
\tablecomments{Electron temperatures and ionic and total abundances for objects with an [\ion{O}{3}] $\lambda4363$ or [\ion{N}{2}] $\lambda5755$ line signal to noise ratio of $3\sigma$ or greater.
Electron temperatures were calculated using either the [\ion{O}{3}] ($\lambda4959 + \lambda5007$)/$\lambda4363$ or the [\ion{N}{2}] ($\lambda6548 + \lambda6584$)/$\lambda5755$ diagnostic line ratio.} 
\end{deluxetable}
				
% Table 8: 
\begin{deluxetable}{lccccc}
\tablenum{4.9}
\tabletypesize{\scriptsize}
\setlength{\tabcolsep}{0.05in} 
\tablecaption{Ionic and Total Abundances for MMT Observations of NGC 628 \label{tbl8}}
\tablewidth{0pt}
\tablehead{}
\startdata			
%{H$\alpha$ Region} 			& {vZ6}  			& {C-4}			& {E-6}			& {F-1}			\\
{H$\alpha$ Region} 				& {+295-16}  		& {-277+240}		& {+186+355}		& {+503+208}		\\	
\hline
t(O$_2$)$_{measured}$ (K)       	& \nodata			& \nodata			& \nodata			& \nodata			\\
t(O$_3$)$_{measured}$ (K)		& 10200$\pm$200	& \nodata			& 10700$\pm$1000	& 11800$\pm$200	\\
t(N$_2$)$_{measured}$ (K)		& \nodata			& \nodata			& \nodata			& \nodata			\\
t(S$_3$)$_{measured}$ (K)		& \nodata			& \nodata			& \nodata			& \nodata			\\
\\
t(O$_2$)$_{used}$ (K)       		& 11200$\pm$500	& 11800$\pm$2300	& 11500$\pm$1100	& 12100$\pm$500	\\
t(O$_3$)$_{used}$ (K)			& 10200$\pm$200	& 12500$\pm$2500	& 10700$\pm$1000	& 11800$\pm$200	\\
t(N$_2$)$_{used}$ (K)			& 11200$\pm$500	& 11800$\pm$2300	& 11500$\pm$1100	& 12100$\pm$500	\\
t(S$_3$)$_{used}$ (K)			& 10100$\pm$500	& 12100$\pm$2400	& 10600$\pm$1000	& 11500$\pm$500	\\
\\
O$^+$/H$^+$ ($\times10^5$)  		& 3.63$\pm$0.68	& 5.86$\pm$4.30	& 5.35$\pm$1.87	& 4.70$\pm$0.71	\\
O$^{++}$/H$^{+}$ ($\times10^5$)  	& 15.7$\pm$1.1	& 3.37$\pm$1.81	& 10.6$\pm$3.0	& 7.23$\pm$0.43	\\
O/H ($\times10^5$)				& 19.4$\pm$1.3	& 9.24$\pm$4.66	& 15.9$\pm$3.6	& 11.9$\pm$0.9	\\
12 + log(O/H)$_{D}$ (dex)		& 8.29$\pm$0.03	& 7.97$\pm$0.18\D	& 8.20$\pm$0.09\D	& 8.08$\pm$0.03	\\
12 + log(O/H)$_{P}$ (dex)		& 7.82$\pm$0.02	& 8.04$\pm$0.02	& 7.94$\pm$0.02	& 7.97$\pm$0.03	\\
\\
N$^{+}$/H$^{+}$ ($\times10^6$)	& 1.66$\pm$0.24	& 2.86$\pm$1.33	& 2.27$\pm$0.52	& 1.86$\pm$0.44	\\
N ICF						& 5.329$\pm$0.076	& 1.575$\pm$0.686	& 2.976$\pm$0.253	& 2.537$\pm$0.092	\\
log(N/O) (dex)					& -1.33$\pm$0.04	& -1.30$\pm$0.05	& -1.37$\pm$0.03	& -1.40$\pm$0.08	\\
N/H ($\times10^6$)				& 9.07$\pm$1.15	& 4.61$\pm$2.39	& 6.80$\pm$1.60	& 4.75$\pm$0.99	\\
12 + log(N/H) (dex)				& 6.96$\pm$0.05	& 6.66$\pm$0.18	& 6.83$\pm$0.09	& 6.68$\pm$0.08	\\
\\
S$^{+}$/H$^{+}$ ($\times10^7$)	& 2.93$\pm$0.43	& 6.30$\pm$0.28	& 4.19$\pm$0.90	& 4.37$\pm$0.46	\\
S$^{++}$/H$^{+}$ ($\times10^7$)	& 31.5$\pm$6.8	& \nodata			& 28.8$\pm$1.34	& 15.9$\pm$6.4	\\
S ICF						& 1.639$\pm$0.170	& \nodata			& 1.305$\pm$0.139	& 1.258$\pm$0.043	\\
S/O							& 0.029$\pm$0.005 	& \nodata			& 0.027$\pm$0.011	& 0.021$\pm$0.006	\\
log(S/O) (dex)					& -1.54$\pm$0.07	& \nodata			& -1.57$\pm$0.15	& -1.67$\pm$0.10	\\
S/H ($\times10^6$)				& 5.64$\pm$0.90	& \nodata			& 4.31$\pm$1.42	& 2.55$\pm$0.65	\\
12 + log(S/H) (dex)				& 6.75$\pm$0.06	& \nodata			& 6.63$\pm$0.12	& 6.41$\pm$0.10	\\
\\
Ne$^{++}$/H$^{+}$ ($\times10^6$)	& 36.3$\pm$3.8	& 6.46$\pm$4.44	& 25.8$\pm$9.35	& 15.6$\pm$1.2	\\
Ne ICF						& 1.231$\pm$0.076	& 2.738$\pm$0.686	& 1.506$\pm$0.253 	& 1.650$\pm$0.092	\\
Ne/O							& 0.23$\pm$0.02	& 0.19$\pm$0.04	& 0.24$\pm$0.02	& 0.22$\pm$0.01	\\
log(Ne/O) (dex)					& -0.64$\pm$0.04	& -0.72$\pm$0.08	& -0.61$\pm$0.04	& -0.67$\pm$0.02	\\
Ne/H ($\times10^5$)				& 4.47$\pm$0.54	& 1.77$\pm$0.96	& 3.88$\pm$0.94	& 2.57$\pm$0.23	\\
12 + log(Ne/H) (dex)				& 7.65$\pm$0.053	& 7.25$\pm$0.19	& 7.59$\pm$0.09	& 7.41$\pm$0.04	\\
\\
Ar$^{++}$/H$^{+}$ ($\times10^7$)	& \nodata			& \nodata			& \nodata			& \nodata			\\
Ar ICF						& \nodata			& \nodata			& \nodata			& \nodata			\\
log(Ar/O) (dex)					& \nodata			& \nodata			& \nodata			& \nodata			\\
Ar/H ($\times10^6$)				& \nodata			& \nodata			& \nodata			& \nodata			\\
12 + log(Ar/H) (dex)				& \nodata			& \nodata			& \nodata			& \nodata			\\
\enddata
\tablecomments{Electron temperatures and ionic and total abundances for objects with an [\ion{O}{3}] $\lambda4363$ or [\ion{N}{2}] $\lambda5755$ line signal to noise ratio of $3\sigma$ or greater.
Electron temperatures were calculated using either the [\ion{O}{3}] ($\lambda4959 + \lambda5007$)/$\lambda4363$ or the [\ion{N}{2}] ($\lambda6548 + \lambda6584$)/$\lambda5755$ diagnostic line ratio.
{\D}Abundance based on a 3$\sigma$ detection of [\ion{O}{3}] $\lambda$4363.} 
\end{deluxetable}

%----------------------------------------------------------------------------------------------------------------------

% Table 9: 
\begin{deluxetable}{lccccccc}
\tabletypesize{\scriptsize}
\tablenum{4.10}
\setlength{\tabcolsep}{0.05in} 
\tablecaption{Ionic and Total Abundances for MMT Observations of NGC 2403 \label{tbl9}}
\tablewidth{0pt}
\tablehead{}
\startdata										
%{H$\alpha$ Region} 			& {35}  			& {24}			& {38}			& {44}			& {9}				& {376}			& {423}			\\	
{H$\alpha$ Region} 				& {-7+36} 			& {-30+45}		& {+13+31}		& {+104+24}		& {-133-146}		& {+376-106}		& {-423-10}		\\	
\hline
t(O$_2$)$_{measured}$ (K)       	& 8500$\pm$500	& 9400$\pm$400	& 8100$\pm$500	& 10100$\pm$200	& \nodata			& \nodata			& \nodata			\\
t(O$_3$)$_{measured}$ (K)       	& 8900$\pm$600	& \nodata			& 9600$\pm$800	& 8700$\pm$200	& 11100$\pm$200	& 9700$\pm$300	& 11300$\pm$100	\\
t(N$_2$)$_{measured}$ (K)       	& 8900$\pm$400	& 8400$\pm$700	& 8300$\pm$300	& 9300$\pm$400	& 10000$\pm$500	&\nodata			& 10800$\pm$1500	\\
t(S$_3$)$_{measured}$ (K)       	& \nodata			& \nodata			& \nodata			& \nodata			& \nodata			& \nodata			& \nodata			\\
\\
t(O$_2$)$_{used}$ (K)       		& 8900$\pm$400	& 8400$\pm$400	& 8300$\pm$300	& 9300$\pm$400	& 10000$\pm$500	& 10900$\pm$500	& 10800$\pm$1500		\\
t(O$_3$)$_{used}$ (K)       		& 8900$\pm$600	& 7700$\pm$700	& 9600$\pm$800	& 8700$\pm$200	& 11100$\pm$200	& 9700$\pm$300	& 11300$\pm$100	\\
t(N$_2$)$_{used}$ (K)       		& 8900$\pm$400	& 8400$\pm$700	& 8300$\pm$300	& 9300$\pm$400	& 10000$\pm$500	& 10900$\pm$500	& 10800$\pm$1500	\\
t(S$_3$)$_{used}$ (K)       		& 9100$\pm$600	& 8100$\pm$700	& 9700$\pm$800	& 8900$\pm$500	& 10900$\pm$500	& 9800$\pm$500	& 11100$\pm$500	\\
\\
O$^+$/H$^+$ ($\times10^5$)  		& 18.6$\pm$3.5	& 17.5$\pm$7.5	& 23.0$\pm$4.6	& 13.0$\pm$2.5	& 9.36$\pm$1.97	& 6.26$\pm$1.2	& 6.10$\pm$3.32	\\
O$^{++}$/H$^{+}$ ($\times10^5$)  	& 7.17$\pm$1.64	& 10.8$\pm$2.9	& 5.39$\pm$1.55	& 12.3$\pm$1.0	& 9.66$\pm$0.49	& 13.4$\pm$1.7	& 9.48$\pm$0.47	\\
O/H ($\times10^5$)				& 25.8$\pm$3.9	& 28.3$\pm$8.0	& 28.4$\pm$4.9	& 25.3$\pm$2.7 	& 19.0$\pm$2.0	& 19.7$\pm$2.0	& 15.6$\pm$3.4	\\
12 + log(O/H)$_{D}$ (dex)		& 8.41$\pm$0.06	& 8.45$\pm$0.11\ds	& 8.45$\pm$0.07\st & 8.40$\pm$0.04\ds 	& 8.28$\pm$0.04	& 8.29$\pm$0.04	& 8.19$\pm$0.08	\\
%12 + log(O/H)$_{M}$ (dex)		& 7.82$\pm$0.02	& 8.61			& 8.52			& 8.52			& 8.52			& 7.97			&				\\
12 + log(O/H)$_{P}$ (dex)		& 8.04$\pm$0.04	& 7.76$\pm$0.04	& 7.88$\pm$0.04	& 7.88$\pm$0.03	& 7.92$\pm$0.03	& 7.89$\pm$0.03	& 7.90$\pm$0.03	\\
\\
N$^{+}$/H$^{+}$ ($\times10^6$)	& 13.9$\pm$1.7	& 14.8$\pm$3.9	& 15.4$\pm$2.1	& 8.59$\pm$1.10	& 3.15$\pm$0.43	& 2.61$\pm$0.33	& 1.97$\pm$0.66	\\
N ICF						& 1.386$\pm$0.204	& 1.615$\pm$0.387	& 1.235$\pm$0.237	& 1.950$\pm$0.147	& 2.032$\pm$0.149	& 3.144$\pm$0.118  & 2.554$\pm$0.303	\\
log(N/O) (dex)					& -1.12$\pm$0.03	& -1.07$\pm$0.04	& -1.17$\pm$0.03	& -1.18$\pm$0.03	& -1.43$\pm$0.03	& -1.38$\pm$0.03	& -1.49$\pm$0.04	\\
N/H ($\times10^6$)				& 19.4$\pm$3.2	& 24.0$\pm$7.1	& 19.1$\pm$3.6	& 16.8$\pm$2.1	& 7.05$\pm$0.89	& 8.17$\pm$1.01	& 5.04$\pm$1.21	\\
12 + log(N/H) (dex)				& 7.29$\pm$0.07	& 7.38$\pm$0.11	& 7.28$\pm$0.07	& 7.23$\pm$0.05	& 6.85$\pm$0.05	& 6.91$\pm$0.05	& 6.70$\pm$0.09	\\
\\
S$^{+}$/H$^{+}$ ($\times10^7$)	& 13.1$\pm$1.6	& 14.3$\pm$3.7	& 15.4$\pm$2.0	& 10.5$\pm$1.3	& 7.40$\pm$1.00	& 4.90$\pm$0.61	& 4.29$\pm$1.41	\\
S$^{++}$/H$^{+}$ ($\times10^7$)	& 27.9$\pm$8.5	& 41.3$\pm$15.2	& 15.9$\pm$6.0	& 46.8$\pm$1.2	& 24.1$\pm$4.4	& 36.9$\pm$9.2	& 18.4$\pm$3.4	\\
S ICF						& 1.070$\pm$0.110	& 1.137$\pm$0.188	& 1.022$\pm$0.108	& 1.195$\pm$0.057	& 1.205$\pm$0.057	& 1.325$\pm$0.079	& 1.260$\pm$0.152	\\
S/O							& 0.018$\pm$0.005 	& 0.022$\pm$0.009	& 0.011$\pm$0.003	& 0.027$\pm$0.006	& 0.020$\pm$0.003	& 0.028$\pm$0.006	& 0.018$\pm$0.005	\\
log(S/O) (dex)					& -1.77$\pm$0.10	& -1.65$\pm$0.15	& -1.95$\pm$0.11	& -1.57$\pm$0.09	& -1.70$\pm$0.07	& -1.55$\pm$0.08	& -1.74$\pm$0.11	\\
S/H ($\times10^6$)				& 4.39$\pm$0.97	& 6.32$\pm$1.89	& 3.20$\pm$0.72	& 6.84$\pm$1.29	& 3.97$\pm$0.49	& 5.54$\pm$0.98	& 2.86$\pm$0.51	\\
12 + log(S/H) (dex)				& 6.64$\pm$0.09	& 6.80$\pm$0.11	& 6.50$\pm$0.09	& 6.84$\pm$0.08	& 6.58$\pm$0.05	& 6.74$\pm$0.07	& 6.46$\pm$0.07	\\
\\
Ne$^{++}$/H$^{+}$ ($\times10^6$)	& 8.59$\pm$2.49	& 15.6$\pm$5.4	& 8.14$\pm$2.97	& 19.1$\pm$1.9	& 17.6$\pm$1.1	& 23.5$\pm$3.7	& 18.8$\pm$1.2	\\
Ne ICF						& 3.591$\pm$0.204	& 2.625$\pm$0.387 	& 5.260$\pm$0.237	& 2.052$\pm$0.147	& 1.969$\pm$0.149	& 1.466$\pm$0.118	& 1.643$\pm$0.303	\\
Ne/O							& 0.12$\pm$0.01	& 0.15$\pm$0.01	& 0.15$\pm$0.01	& 0.15$\pm$0.01	& 0.18$\pm$0.01	& 0.18$\pm$0.01	& 0.20$\pm$0.01	\\
log(Ne/O) (dex)					& -0.92$\pm$0.04	& -0.84$\pm$0.04	& -0.82$\pm$0.04	& -0.81$\pm$0.03	& -0.74$\pm$0.02	& -0.76$\pm$0.03	& -0.70$\pm$0.02	\\
Ne/H ($\times10^5$)				& 3.09$\pm$0.53	& 4.10$\pm$1.24	& 4.28$\pm$0.84	& 3.92$\pm$0.48	& 3.46$\pm$0.42	& 34.5$\pm$4.3	& 3.09$\pm$0.69	\\
12 + log(Ne/H) (dex)				& 7.49$\pm$0.07	& 7.61$\pm$0.11	& 7.63$\pm$0.08	& 7.59$\pm$0.05	& 7.54$\pm$0.05	& 7.54$\pm$0.05	& 7.50$\pm$0.09	\\
\\
Ar$^{++}$/H$^{+}$ ($\times10^7$)	& 10.5$\pm$2.0	& 12.2$\pm$2.5	& 6.02$\pm$1.32	& 11.5$\pm$1.8	& \nodata			& \nodata			& \nodata			\\
Ar ICF						& 2.000$\pm$0.993	& 1.619$\pm$0.730	& 2.825$\pm$0.183	& 1.468$\pm$0.078	& \nodata			& \nodata			& \nodata			\\
log(Ar/O) (dex)					& -2.09$\pm$0.11	& -2.16$\pm$0.14	& -2.22$\pm$0.12	& -2.17$\pm$0.09	& \nodata			& \nodata			& \nodata			\\
Ar/H ($\times10^6$)				& 2.10$\pm$0.47	& 1.98$\pm$0.48	& 1.70$\pm$0.45	& 1.69$\pm$0.32	& \nodata			& \nodata			& \nodata			\\
12 + log(Ar/H) (dex)				& 6.32$\pm$0.09	& 6.30$\pm$0.09	& 6.23$\pm$0.10	& 6.23$\pm$0.07	& \nodata			& \nodata			& \nodata			\\

\enddata
\tablecomments{Electron temperatures and ionic and total abundances for objects with an [\ion{O}{3}] $\lambda4363$ or [\ion{N}{2}] $\lambda5755$ line signal to noise ratio of $4\sigma$ or greater.
Electron temperatures were calculated using either the [\ion{O}{3}] ($\lambda4959 + \lambda5007$)/$\lambda4363$ or the [\ion{N}{2}] ($\lambda6548 + \lambda6584$)/$\lambda5755$ diagnostic line ratio. \\
{\D}Abundance based on a 3$\sigma$ detection of [\ion{O}{3}] $\lambda$4363. \\
{\st}\ion{H}{2} region overlaps with the G97 sample.} 
\end{deluxetable}

\end{document}